\documentclass[12pt]{article}
\usepackage{latexsym}

\csname @addtoreset\endcsname{equation}{section}
\newcommand{\Ka}{K\"ahler}
\newcommand  {\Rbar} {{\mbox{\rm$\mbox{I}\!\mbox{R}$}}}
\def\Im{{\rm Im ~}}
\newsavebox{\uuunit}
\sbox{\uuunit}
    {\setlength{\unitlength}{0.825em}
     \begin{picture}(0.6,0.7)
        \thinlines
        \put(0,0){\line(1,0){0.5}}
        \put(0.15,0){\line(0,1){0.7}}
        \put(0.35,0){\line(0,1){0.8}}
       \multiput(0.3,0.8)(-0.04,-0.02){12}{\rule{0.5pt}{0.5pt}}
     \end {picture}}
\newcommand {\unity}{\mathord{\!\usebox{\uuunit}}}
\newcommand{\Poin}{Poincar\'e}

\def\<{\langle}
\def\>{\rangle}

\def\bpl{\Biggl(}
\def\bpr{\Biggr)}

\def\R{\cal{R}}
\def\a{\alpha}
\def\b{\beta}
\def\d{\delta}

\def\D{{\cal D}}
\def\e{\epsilon}
\def\ve{\varepsilon}

\def\F{{\cal F}}

\def\g{\gamma}
\def\l{\lambda}

\def\s{\sigma}
\def\t{\theta}

\def\w{\omega}
\def\O{\Omega}
\def\zb{\bar z}

\def\be{\bar{\epsilon}}

\def\der{\partial}
\def\bq{\begin{equation}}
\def\eq{\end{equation}}
\def\brr{\begin{eqnarray}}
\def\err{\end{eqnarray}}
\def\ba{\left(\begin{array}}
\def\ea{\end{array}\right)}

\newcommand{\nslash}{/\!\!\!\!\nabla}

\newcommand{\Aslash}{/\!\!\!\! A}

\newcommand{\Qslash}{/\!\!\!\! {\cal Q}}

\def\dslash{\hbox{\ooalign{$\displaystyle\partial$\cr$/$}}}
\def\Dslash{\hbox{\ooalign{$\displaystyle D$\cr$\hspace{.03in}/$}}}

\def\ba{\left(\begin{array}}
\def\ea{\end{array}\right)}

\newcommand\eqn[1]{(\ref{#1})}
\newcommand{\rmi}{{\rm i}}
\begin{document}
\newcommand\ft[2]{{\textstyle\frac{#1}{#2}}}
\begin{titlepage}
\begin{raggedleft}
KUL-TF-99/08\\
{\tt hep-th/9904066}\\[.5cm]
April 1999\\
\end{raggedleft}
\ \vskip2cm
\begin{center}
{\bf \Large A symplectic covariant formulation of special\\[8pt]
\Ka\ geometry in  superconformal calculus.
}
\vskip 10.3mm
{\bf Piet~Claus, Kor~Van Hoof and Antoine~Van Proeyen$^*$}
\vskip2cm
Instituut voor Theoretische Fysica, Katholieke
Universiteit Leuven,\\ Celestijnenlaan 200D, B-3001 Leuven, Belgium\\
\end{center}
\vfill
\begin{center}
{\bf Abstract}
\end{center}
\begin{quote}
We present a formulation of the coupling of vector multiplets to
$N=2$ supergravity which is symplectic covariant (and thus is not based on a prepotential)
and uses superconformal tensor calculus.
 We do not start from an action, but
from the combination of the generalized Bianchi identities of the
vector multiplets in superspace, a symplectic definition of special
K\"ahler geometry, and the supersymmetric partners of the
corresponding constraints. These involve the breaking to super-\Poin\
symmetry, and lead to on-shell vector multiplets.\\
This symplectic approach gives the framework to formulate vector
multiplet couplings
using a weaker defining constraint for special \Ka\ geometry, which is an extension
of older definitions of special \Ka\ manifolds for some cases with
only one vector multiplet.
\vskip 2mm
 \hrule width 5.cm
\vskip 2.mm
{\small
\noindent $^*$ Onderzoeksdirecteur FWO, Belgium }
\end{quote}
\end{titlepage}
\section{Introduction}
The understanding of the most general coupling of vector multiplets
in $N=2$ supersymmetry or supergravity is important in very
different contexts. It is used in the context of the low-energy
effective actions of supersymmetric gauge theories and their
coupling to hypermultiplets \cite{SW}. It is also an essential
element in the compactification of string theory on Calabi--Yau
manifolds \cite{CY}. This general coupling has been studied for the
supergravity case in \cite{dWLPQVP,dWVP} and has been given the name
special geometry \cite{strom}. The similar coupling in rigid
supersymmetry was obtained in \cite{STG} and is referred to as `rigid
special geometry'.
\par
Historically, the coupling of several $N=2$ matter multiplets to $N=2$
supergravity in four dimensions was found using superconformal tensor calculus
\cite{dWVHVP, dRvHdWVP, dWLPQVP, dWVP, dWLVP}. Superconformal actions are
built with representations of a larger algebra, the $N=2$ superconformal one.
The residual symmetries are broken by introducing two compensating multiplets
(one vector multiplet to give rise to the graviphoton in the Poincar\'e
gravity multiplet and a hyper-, a linear or a nonlinear multiplet) to end up
with an $N=2$ Poincar\'e supergravity theory coupled to $N=2$ matter
multiplets. Reviews on superconformal tensor calculus can be found in
\cite{sctc}.
The superconformal construction clarifies the origin of many terms
in the action.
Here we will confine ourselves to the coupling of vector
multiplets to supergravity, where after breaking the superconformal symmetry,
the complex scalars of the vector multiplets form a special \Ka\ manifold.
\par
A prepotential, a holomorphic function of second order, was an essential
ingredient to construct the theory. In \cite{CDFF,FL,ABCDFF}, other approaches were used to describe the coupling of vector
multiplets to supergravity.
\par
Electric--magnetic duality transformations in four dimensions manifest
themselves by symplectic transformations \cite{dual}.
Symplectic transformations in a special \Ka\ manifold have
been studied in \cite{dWVP,CecFerGir}.  The duality symmetry is not a
symmetry of the complete action, but only of the equations of motion. In
particular, the prepotential is not an invariant of the symplectic
transformations. On the other hand, for a coordinate-free formulation of
special geometry \cite{CDFF}, the symplectic symmetry is an essential
ingredient.
The symplectic set-up also clarified the link to
Calabi--Yau manifolds \cite{CY}. In \cite{CDFVP}, vector multiplet
couplings to supergravity were constructed for which no prepotential existed,
by performing a symplectic transformation of an action based on a prepotential.
The resulting action was thus not based on a prepotential.
The first and main purpose of this paper is to obtain a symplectic covariant
formulation of the coupling of vector multiplets to $N=2$ supergravity which
at the same time uses superconformal tensor calculus. In particular, it should
thus contain the coupling of \cite{CDFVP}. For obtaining
an action in superconformal tensor calculus, one needs a prepotential, and
hence to give up the symplectic covariance. The combination of superconformal and
symplectic covariance will, however, be possible if we only construct
equations of motions without an action.
\par
The various possible actions and geometric formulations were
compared in \cite{CRTVP}, and one has arrived at a new definition of
special geometry (also the definition for the case of rigid
supersymmetry is given there). Remarkably, it was also noticed that
one part of the definition, expressed by differential constraints,
can be formulated in two different ways. These two forms are
equivalent when more than one vector multiplet is coupled to
supergravity, but inequivalent if only one vector multiplet is
coupled. The presentation in \cite{CRTVP} contained the constraints
such that for one vector multiplet the coupling known previously
(e.g.\ from superconformal tensor calculus) is obtained.
However, it was noted that another form of the constraints is possible,
which is also symplectic covariant. Obvious physical arguments could not
exclude the existence of hitherto unknown couplings of 1 vector
multiplet to supergravity which obey the weaker constraint, and not
the stronger one. The second aim of this paper is to show that such
new couplings are indeed possible.
\par
To be more explicit, let us repeat here one of the formulations of the
3 equivalent definitions of a special \Ka\ manifold of \cite{CRTVP}\footnote{The
first one is not explicitly symplectic covariant, but we could as well have
discussed here definition~2, where the constraint relevant for the discussion
below was formulated as $\langle v , \partial_{\alpha} v \rangle =0$. The
alternative form is then $\langle \partial_\alpha v , \partial_\beta v
\rangle =0$.}.
The most suitable definition
for our discussion is formulated in terms of symplectic products. It is the one
 which was denoted as definition~3.
\vspace{2mm}
\par
\noindent{\it Definition of a special \Ka\ manifold}\vspace{2mm}
\par
Take a complex manifold ${\cal M}$. Suppose we have in every chart a
$2(n+1)$-component
vector $V(z^{\alpha},\bar z^{\alpha})$ such that on overlap regions there are
transition functions
 of the form
\begin{equation}
e^{\frac{1}{2}(f(z^{\alpha})
-\bar f(\bar z^{\alpha}))}\,S\,,
\end{equation}
with $f$ a holomorphic function and
$S$ a constant $Sp(2(n+1),\Rbar)$  matrix. (These transition functions
have to satisfy the cocycle condition.) Take a $U(1)$ connection of the form
$\kappa_{\alpha}\,dz^{\alpha}+
\kappa_{\bar \alpha}\,d\bar z^{\alpha}$ with
\begin{equation}
\kappa_{\bar\alpha}= -\overline{\kappa_{\alpha}}\,,  \label{defkappabar}
\end{equation}
under which $\bar V$ has opposite
weight as $V$.
Denote the covariant derivative by ${\cal D}$:
\begin{equation}
\begin{array}{ll}
U_\alpha\equiv {\cal D}_{\alpha} V\equiv
\partial_{\alpha} V+\kappa_{\alpha} V
\,, \qquad&
{\cal D}_{\bar \alpha}V\equiv
\partial_{\bar\alpha} V+\kappa_{\bar\alpha} V\,,\\
\bar U_{\bar \alpha}\equiv {\cal D}_{\bar\alpha}\bar V\equiv
\partial_{\bar\alpha}\bar V-\kappa_{\bar\alpha}\bar V
\,,\qquad&
{\cal D}_{\alpha}\bar V\equiv
\partial_{\alpha}\bar V-\kappa_{\alpha}\bar V \,.
\end{array}            \label{defD}
\end{equation}
We impose the following conditions:
\begin{eqnarray}
&1.& \langle V,\bar V\rangle =\rmi\,,    \label{VVbar} \\
&2.& {\cal D}_{\bar \alpha}V= 0\,,    \label{DbarV} \\
&3.& {\cal D}_{[\alpha}U_{\beta]}=0\,,   \label{DU=0}\\
&4.& \langle V,U_{\alpha}\rangle =0 \,, \label{VUal}
\end{eqnarray}
where $\langle \cdot ,\cdot \rangle$ denotes the symplectic inner
product, e.g.\ $\langle V,\bar V\rangle= V^T\Omega\bar V$, with an
antisymmetric matrix $\Omega$, which has as standard form
\begin{equation}
\Omega_{st}=\pmatrix{0&\unity \cr -\unity &0}\,.\label{standardOmega}
\end{equation}
Define
\begin{equation}  \label{defg}
 g_{\alpha\bar\beta}\equiv \rmi\langle U_{\alpha},\bar U_{\bar\beta}\rangle \,,
\end{equation}
where $\bar U_{\bar\beta}$ denotes the complex conjugate of $U_{\alpha}$.
If this is a positive-definite metric,
${\cal M}$ is called a special \Ka\ manifold.
\par
It can then be shown that locally a function $K'$ exists such that
\begin{equation}
 \label{kappa}
 \kappa_{\alpha}=\ft{1}{2}\partial_\alpha K' \,, \qquad
 \kappa_{\bar\alpha}=-\overline{\kappa_{\alpha}}=
 -\ft{1}{2}\partial_{\bar\alpha}\bar K'\,.
\end{equation}
The real part of $K'$ is the \Ka\ potential $K$. If there is an imaginary part
$\Im K'$, then
\begin{equation}
V'=e^{\rmi\Im K'/2}V \,,\label{V'V}
\end{equation}
satisfies the same constraints with $K'$ replaced by the real $K$.
\par
As discussed at the end of section~4.2.2 in \cite{CRTVP}, the constraints
have a clear physical interpretation, related to the positivity of kinetic terms
in the action. However, as suggested there already, the
fourth constraint \eqn{VUal} could be replaced by
\begin{equation}
4'.\ \ \langle U_\alpha, U_\beta\rangle =0 \,, \label{UU}
\end{equation}
without violating the physical arguments. The constraint $4$ implies $4'$, by
taking a covariant derivative and antisymmetrizing, and with $4'$ it was shown
that $4$ follows when $n>1$. However, for $n=1$, equation $4'$ is empty.
Taking $4'$ as constraint thus allows $\langle V, U_z\rangle \neq 0$.
Such $N=2$ models would be new, and this possibility will
be investigated in this paper. It will be called `the special case'.
The case with $n>1$ or $\langle V, U_\a \rangle =0$ will be called
`the generic case'.
\par
In appendix~C of \cite{CRTVP}, two $n=1$ examples are given where the condition
\eqn{VUal}
is not fulfilled. In these examples it was shown that the relaxation
of that constraint leads to models not allowed by other definitions
of special geometry. Here we will first give further evidence of the
non-triviality of this condition. The main result will be that indeed
models which violate \eqn{VUal}, still allow an $N=2$ supersymmetric formulation.
\par
The scalars of the special \Ka\ manifolds are contained in the $\theta=0$
sector of chiral superfields of $N=2$ supersymmetry.   A
chiral multiplet is a reducible representation of $N=2$ supersymmetry.
After imposing suitable constraints, it gives a vector multiplet.
In rigid supersymmetry these constraints can be written in superspace
for a symplectic section of superfields or, in components, as a linear
multiplet of constraints of
symplectic sections \cite{CRTVP,LW}.
With a standard symplectic metric \eqn{standardOmega}, the
rigid special \Ka\ constraints can be used to write the lower part
of the symplectic sections $V$ in terms of the upper one.
The reducibility constraints for the lower parts of the sections
then give rise to the field equations of the fields in the upper parts.
We want to use this approach to construct the field equations of
vector multiplets, coupled to $N=2$ supergravity.
\par
Chiral and vector multiplets can also be defined as representations of the local
superconformal algebra. Then the fields of the gauge multiplet of the
superconformal gauge
invariances, which is the Weyl multiplet, enter in the transformation rules
of the multiplets \cite{dWVHVP, dRvHdWVP}.
To describe the coupling of $n$ on-shell vector multiplets to
supergravity, we will start from $2n+2$ chiral multiplets.
The linear multiplet of constraints which reduce these chiral multiplets
to vector multiplets in supergravity will contain additional terms with fields
of the Weyl multiplet \cite{dWVHVP}.
\par
The equations that follow by supersymmetry from this weak definition of
special \Ka\ geometry are derived for the complete set of $2n+2$ chiral
multiplets. The constraints defining special \Ka\ geometry involve a
breaking of dilatations and the $U(1)$ transformations in the
superconformal group. We also choose a symplectic fermionic
constraint as the gauge choice for $S$-supersymmetry. Special conformal
symmetry is broken by a choice for the dilatation gauge field as in
previous approaches. So finally, this leads to the breaking of superconformal
to super-Poincar\'e spacetime symmetry with a residual internal $SU(2)$ in a
consistent way, without relying on a prepotential or an action.
Combining the reducibility constraints with the constraints of
special \Ka\ geometry we find $n$ on-shell vector multiplets,
coupled to $24+24$ supergravity components, remnants of the Weyl
multiplet.
\par
These $24+24$ components reside in a `current multiplet', which we
identify as a reduced chiral self-dual superfield. The full
supergravity equations, however, would rely on a second compensating
multiplet, which is independent of the symplectic formulation. For
these aspects we refer to the 3 known  constructions of auxiliary
field formulations \cite{auxf,dWVHVP}.
\par
In section~\ref{s:bb}, the building blocks of the construction, the
Weyl multiplet and the chiral multiplet, are given. Their supersymmetry transformation
rules and the constraint to make a vector multiplet out of a chiral are
recapitulated. In section~\ref{s:kahler} the special K\"ahler constraints and the supersymmetric
relatives are treated for the most general case. In section~\ref{s:gBI} we
combine the constraints imposed on the chiral multiplets and those found in
section~\ref{s:kahler} to find  on-shell vector multiplets. Finally,
we comment on the remaining off-shell components of supergravity and
their field equations. We recapitulate our results in
section~\ref{ss:concl}.
\section{The building blocks of the construction}
\label{s:bb}
In this section we review the Weyl multiplet, i.e.\ the gauge multiplet of the
$N=2$ superconformal symmetry, and the superconformal chiral multiplet,
coupled to the Weyl multiplet. Most of the material presented here is
well known (see e.g., \cite{dRvHdWVP, vecten})\footnote{However, here we use different
normalizations, more suited for a manifestly symplectic formulation of the
theory. We use the notations of \cite{trsummer}. So the old supersymmetry
parameters are $1/\sqrt{2}$ the new ones and the old fermionic fields
are $\sqrt{2}$ the new ones. Also keep in mind that $\ve^{0123} = \rmi$.}.

\subsection{The Weyl multiplet}\label{s:Weyl}

The Weyl multiplet is the gravitational multiplet of $N=2$ superconformal
gravity. It contains the gauge fields $e^a_\mu, \omega^{ab}_\mu, b_\mu,
f_\mu^a, {\cal V}_\mu^i{}_j, A_\mu, \psi^i_\mu$ and $\phi^i_\mu$. They are,
respectively, gauge fields of general coordinate transformations, Lorentz
rotations, dilatations, special conformal boosts, chiral $SU(2)$ and $U(1)$,
supersymmetry and special supersymmetry. The representation is
completed by the Lorentz tensor $T^{ij}_{ab}$, antisymmetric in $[ij]$,
the spinor $\chi^i$ and the
scalar $D$. Note that $T^{ij}_{ab}$ is the antiself-dual tensor, and
its complex conjugate $T_{abij}$ is self-dual.
The spin connection and the gauge fields for the special conformal
transformations and special supersymmetry are composite gauge fields, given by
\brr \w_\mu^{ab} &=& -2e^{\nu[a}\der_{[\mu}e_{\nu]}{}^{b]}
     -e^{\nu[a}e^{b]\s}e_{\mu c}\der_\s e_\nu{}^c
     -2e_\mu{}^{[a}e^{b]\nu}b_\nu   \nonumber\\
     & & -\ft{1}{2}(2\bar{\psi}_\mu^i\g^{[a}\psi_i^{b]}
     +\bar{\psi}^{ai}\g_\mu\psi^b_i+{\rm h.c.})\,, \nonumber\\
     \phi_\mu^i &=& (\s^{\rho\s}\g_\mu-\ft{1}{3}\g_\mu\s^{\rho\s})
     \left({\cal D}_\rho\psi_\s^i-\ft{1}{8}\s\cdot T^{ij}\g_\rho\psi_{\s j}\right)
     +\ft{1}{2}\gamma _\mu \chi^i\,, \nonumber\\
     f_\mu{}^a &=& \ft12 {\cal R}_\mu^a -\ft12e_\mu{}^a f_\nu{}^\nu
     -\rmi\ft14 e^{a\nu}\ve_{\mu\nu}{}^{\rho\s}{\hat R}_{\rho\s}(U(1))
     +\ft1{16}T_{ij}^{ab}T_{\mu b}^{ij} -\ft34 e_\mu{}^a D\label{dependent} \\
     &&+ \Big({\bar \psi}^i_{[\mu}\s^{ab}\phi_{\nu]i}
     + \ft12{\bar\psi}^i_{[\mu}T^{ab}_{ij}\psi^j_{\nu]} -\ft32 {\bar
     \psi}^i_{[\mu}\g_{\nu]}\s^{ab}\chi_i
     - {\bar\psi}^i_{[\mu}\g_{\nu]}{\hat R}^{ab}({\rm Q})_i
     + {\rm h.c.}\Big)e_b{}^\nu \,.\nonumber
\err
The following expressions are used in $f_\mu{}^a$:
\brr
     f_\mu{}^\mu &=& \ft{1}{6}{\cal R}-D
     -\Big(\ft{1}{6}e^{-1}\ve^{\mu\nu\rho\s}\bar{\psi}_\mu^i\g_\nu
     {\cal D}_\rho\psi_{\s i}
     -\ft{1}{6}\bar{\psi}_\mu^i\psi_\nu^j\,T_{ij}^{\mu\nu}
     -\ft{1}{2}\bar{\psi}_\mu^i\g^\mu\chi_i+{\rm h.c.}\Big)\,,\nonumber \\
     {\hat R}_{\rho\s}(U(1)) & = & 2\partial_{[\mu}A_{\nu]} -\rmi\left(2{\bar
     \psi}^i_{[\mu}\phi_{\nu]i} +\ft32 {\bar \psi}^i_{[\mu}\g_{\nu]}\chi_i
     +{\rm h.c.}\right)\,,\nonumber \\
     {\hat R}^{ab}( Q)^i & = & 2{\cal D}_{[\mu} \psi_{\nu]}^i
     -\g_{[\mu}\phi_{\nu]}^i-\ft14 \s\cdot T^{ij}\g_{[\mu}\psi_{\nu] j}
     \,.
\label{dependent2}
\err
Also, ${\cal D}_\mu$ is covariant with respect to the linearly
realized symmetries: Lorentz transformations, dilatations, $U(1)$ and
$SU(2)$, i.e.
\begin{equation}
{\cal D}_\mu \psi_\nu^i= \left( \partial_\mu
-\ft12\omega_{\mu}^{ab}\sigma_{ab} +\ft12 b_\mu +\ft 12\rmi A_\mu \right) \psi_\nu^i
+\ft12 {\cal V}_\mu{}^i{}_j\psi_\nu^j\ .
\end{equation}
Furthermore,
${\cal R} = e^\mu_a e^\nu_b{\cal R}_{\mu\nu}^{ab}$ is the Ricci scalar derived
from the Riemann tensor
\bq
{\cal R}_{\mu\nu}{}^{ab} = 2\partial_{[\mu}\omega_{\nu ]}^{ab} -
2\omega_{[\mu}^{ac}\omega_{\nu ]c}{}^b
\eq
and
\bq
{\cal R}^a_\mu = e^\nu_b{\cal R}_{\mu\nu}^{ab}\ .
\eq
The transformation rules of the independent fields of the Weyl
multiplet under supersymmetry, special supersymmetry and special conformal
transformations (with parameters $\e^i$, $\eta^i$ and $\Lambda^a_K$)
are
\brr \d e_\mu{}^a &=&
     \bar{\e}^i\g^a\psi_{\mu i}+{\rm h.c.}\,, \nonumber\\
      \d\psi_\mu^i &=& {\cal D}_\mu\e^i
      -\ft18 \s\cdot T^{ij}\g_\mu\e_j
      -\ft12 \g_\mu\eta^i \,,\nonumber\\
      \d b_\mu &=&
      \ft12\bar{\e}^i\phi_{\mu i}
      -\ft34\bar{\e}^i\g_\mu\chi_i
      -\ft12\bar{\eta}^i\psi_{\mu i}+{\rm h.c.} +\Lambda_K^a\,e_{\mu\,a}
      \,,\nonumber\\
      \d A_\mu &=& \ft{1}{2}\rmi \bar{\e}^i\phi_{\mu i}
      +\ft{3}{4}\rmi\bar{\e}^i\g_\mu\chi_i
      +\ft{1}{2}\rmi \bar{\eta}^i\psi_{\mu i}+{\rm h.c.}\,, \nonumber\\
      \d {\cal V}_\mu^i{}_j &=&
      2\bar{\e}_j\phi_\mu^i-3\bar{\e}_j\g_\mu\chi^i+2\bar{\eta}_j\psi_\mu^i
     -({\rm h.c.} \, ; \, {\rm traceless})\,,
      \nonumber\\
      \d T_{ab}^{ ij} &=&
      8 \bar{\e}^{[ i} \hat{R}_{ab}(Q)^{j]}\,, \nonumber\\
      \d\chi^i &=& -\ft1{12}\s\cdot\Dslash T^{ij}\e_j
      +\ft{1}{6}\hat{R}(SU(2))^i{}_{j}\cdot\s\e^j
      -\ft{1}{3}\rmi \hat{R}(U(1))\cdot\s\e^i \nonumber\\
      && +\ft12D\,\e^i
      +\ft1{12}\s\cdot T^{ij}\eta_j \,,\nonumber\\
      \d D &=& \bar\e^i\Dslash\chi_i+{\rm h.c.}\,.
\label{transfo4}\err
The expression for the superconformal covariant curvatures $\hat R$
and the other transformation rules (in terms of the old conventions) were given
in \cite{dWVHVP}.

\subsection{The chiral multiplet}

A chiral multiplet is a reducible representation of the superconformal algebra
\cite{dRvHdWVP}. By imposing a
linear multiplet of constraints it becomes a vector multiplet.
This is an irreducible representation of the superconformal algebra.
The constraints are called the generalized Bianchi identities, because they contain a
Bianchi identity for the tensor in the chiral multiplet.
\par
Later we want to couple vector multiplets to conformal supergravity. The scalars
of these vector multiplets
form a symplectic section. They are the lowest components of a multiplet.
Therefore, all the components of the multiplets have to form such a
symplectic section. This is the reason to start from a $(2n+2)$-dimensional
section of chiral multiplets:
\brr
\tilde \Phi &=& V + \bar \t^i \tilde \O_i + \ft14 \bar \t^i \t^j \tilde
Y_{ij} + \ft14 \ve_{ij} \bar \t^i \s\cdot \tilde {\cal F}^- \t^j\nonumber\\
& &+ \ft16 \ve_{ij} (\bar \t^i \s_{ab} \t^j) \bar \t^k \s^{ab} \tilde \Lambda_k
+ \ft1{48} (\ve_{ij} \bar \t^i \s_{ab} \t^j)^2 \tilde C\,. \label{chiralsup}
\err
The components of the section are denoted by
\bq
\tilde \Phi =\ba{c} \phi^I\\\phi_{F,I} \ea\,,
\eq
with $I=0,\dots,n$, which gives for the components of the chiral superfield
\brr
V&=&\ba{c}X^I\\F_I\ea\,, \hspace{.5in} \tilde \O_i = \ba{c}\O_i^I \\
\O_{F,Ii}\ea\,, \hspace{.5in} \tilde Y_{ij} = \ba{c}Y_{ij}^I\\Y_{F,I ij}\ea\,,
\nonumber\\
\tilde {\cal F}^-_{ab} &=& \ba{c} {\cal F}_{ab}^{I-} \\ {\cal G}_{F,I
ab}^-\ea\,, \hspace{.34in} \tilde \Lambda_i =\ba{c} \Lambda^I_i \\
\Lambda_{F,Ii}\ea\,,
\hspace{.57in} \tilde C
= \ba{c} C^I\\C_{F,I}\ea\,.
\err
This section of multiplets is independent of the existence of a
prepotential. The multiplets starting with $F_I$ are on an equal footing with the
ones starting with $X^I$. As long as we do not impose the constraints to obtain
a vector superfield or the
special K\"ahler constraints these are $2n+2$ independent chiral multiplets.
The full superconformal transformation rules are given by
\brr
\d V &=&  \bar \e^i \tilde \O_i + (\Lambda_D - \rmi \Lambda_A)V\,,\nonumber\\
\d \tilde \O_i &=& \Dslash V \e_i + \ft12 \tilde Y_{ij} \e^j + \ft12
                   \s\cdot \tilde {\cal F}^- \ve_{ij} \e^j + V\eta_i + (\ft32
           \Lambda_D - \ft 12\rmi
                   \Lambda_A) \tilde \O_i + \Lambda_{SU(2)\, i}{}^j \tilde
                   \O_j\,,\nonumber\\
\d \tilde Y_{ij} &=& 2 \be_{(i} \Dslash \tilde \O_{j)} - 2\be^k
                     \tilde\Lambda_{(i} \ve_{j)k} + 2 \Lambda_D \tilde Y_{ij}
                     +2 \Lambda_{SU(2)\, (i}{}^k \tilde Y_{j)k}\,,\nonumber\\
\d \tilde {\cal F}_{ab}^-&=&\ve^{ij} \be_i \Dslash \s_{ab} \tilde \O_j
                            + \be^i \s_{ab} \tilde \Lambda_i -2 \ve^{ij} \bar
                \eta_i \s_{ab}
                            \tilde \O_j + 2 \Lambda_D \tilde {\cal F}^-_{ab}
                \,,\nonumber\\
\d \tilde \Lambda_i &=&-\ft12 \s \cdot \tilde \F^- \stackrel{\leftarrow}
{\Dslash}
                        \e_i - \ft12 \Dslash \tilde Y_{ij} \e_k \ve^{jk} + \ft12
                        \tilde C \e^j \ve_{ij}\nonumber\\
                    & & -\ft18 \Dslash (V \ve^{jk}T_{jk} \cdot \s)\e_i
            - \ft32(\bar
                        \chi_{[i}\g_a \tilde \O_{j]}) \g^a \e_k \ve^{jk}
            \nonumber\\
                    & & - \tilde Y_{ij} \ve^{jk} \eta_k + \ft52 \Lambda_D
                        \tilde \Lambda_i + \ft 12\rmi\Lambda_A \tilde
                        \Lambda_i + \Lambda_{SU(2)\, i}{}^{j}
            \tilde
                        \Lambda_j \,,\nonumber\\
\d \tilde C &=& - 2 \ve^{ij} \be_i \Dslash \tilde \Lambda_j
                -  6 \be_i \chi_j \tilde Y_{kl} \ve^{ik}\ve^{jl}
                +  \ft12 \be_i \s\cdot T_{jk}
        \Dslash \tilde \O_l \ve^{ij}\ve^{kl}
                + 2 \ve^{ij} \bar\eta_i \tilde \Lambda_j\nonumber\\
            & & + 3 \Lambda_D \tilde C + i \Lambda_A \tilde C\,.
\label{transfo}
\err
This superconformal chiral superfield can be reduced to a vector superfield
with the constraints
\brr
0&=&\tilde Y_{ij} - \ve_{ik}\ve_{jl} \tilde Y^{kl}\,,\label{bian1}\\
0&=&\Dslash \tilde \O^i - \ve^{ij} \tilde \Lambda_j\,,\label{bian2}\\
0&=& D^a(\tilde {\F}_{ab}^+ - \tilde {\F}_{ab}^- + \ft14 V
T_{ab\, ij}\ve^{ij} -
\ft14 \bar V T^{ij}_{ab}\ve_{ij})
- \ft32 ( \ve^{ij} \bar \chi_i\g_b \tilde \O_j
- \mbox{h.c.})\,,\label{bian3}\\
0&=&-2 \Box \bar V - \ft 14 \tilde {\cal F}^+_{\mu\nu}
T^{\mu\nu}_{ij}\ve^{ij} - 6 \bar \chi_i \tilde
\O^i - \tilde C\,.
\label{bian4}
\err
The symplectic vector of chiral multiplets with these constraints
define $2n+2$ vector multiplets in
superconformal gravity. The special K\"ahler constraints will relate
them such that one ends up with $n+1$ vectors and $n$ complex scalars
and spinors obeying field equations.
\setcounter{equation}{0}
\section{Gauge choices and special K\"ahler constraints\label{s:kahler}}
To obtain a Poincar\'e supergravity theory of $n$ vector multiplets,
we start from the assumption that the components in the symplectic
sections $V$ are the lowest components of reduced chiral multiplets, as is
the case in previous constructions of matter couplings in $N=2$
supergravity.
To achieve that, we have to impose
the reducibility constraints \eqn{bian1}--(\ref{bian4}) on the chiral multiplets and
suitable constraints that impose restrictions on the sections such that the
resulting theory contains
$n$ physical vector multiplets and the gravity multiplet.
The superfluous symmetries of the superconformal construction need to be broken
by suitable gauge choices.
The symplectic section $V$ can be seen as a function of $n$ scalars $z^\a$ and
their complex conjugates $\bar z^{\bar\a}$ ($\a =1,..., n$). These scalars can
be interpreted as the coordinates of a special K\"ahler manifold.
\par
Having introduced $K'$ in \eqn{kappa}, we have exhausted constraint
\eqn{DU=0}. The remaining relevant constraints are then \eqn{VVbar},
\eqn{DbarV},
and we will take the formulation with \eqn{UU}.
Condition (\ref{VVbar}) gauge fixes the dilatations, choosing
the canonical
kinetic term for the graviton. Equation~(\ref{DbarV}) imposes the
holomorphicity of the scalar fields. For the
symmetry of the kinetic matrix of the vectors, one needs another
constraint, which is \eqn{UU}. In all previous papers on
special geometry, one imposed instead \eqn{VUal}, which is equivalent for
$n>1$, but not for $n=1$ as mentioned in the introduction.
There is no physical argument known to demand \eqn{VUal}, but
up to now, no physical
applications have been found that do not fulfil it.
\par
We have thus seen that
we can look upon equations \eqn{VVbar} and \eqn{DbarV} in two ways.  They
are the defining equations of special geometry, as well, they can be considered
 as gauge choices for the dilatations and chiral $U(1)$ transformations present
in the superconformal algebra.  As we will see below a supersymmetric
extension of these constraints will include the gauge choice of $S$-supersymmetry.
\par
{}From (\ref{defD}) follows
\bq
g_{\a \bar \b}\equiv \der_\a \der_{\bar \b} K =
\rmi \<  U_\a, \bar U_{\bar \b}\>\,.
\eq
Furthermore, we impose the \lq physical' condition (positivity of the
kinetic energy terms of the vectors \cite{CRTVP}) that\footnote{
Keep in mind that for $n>1$ one always has $Z_z = 0$.}
\begin{eqnarray}
\det g_{\a\bar\b} > 0 \ &\qquad\mbox{if }& \<  V,U_\a\> = 0\,, \nonumber\\
g'_{z\bar z} \equiv  g_{z{\bar z}}- Z_z\bar Z_{\bar z}\ > 0
&\qquad\mbox{for }&  Z_z \equiv \<  V, U_z\> \,.
\label{physicalcond}
\end{eqnarray}
Using the constraints it can be shown that
\bq
{\cal W}=(V,U_\a, \bar V, \bar U_{\bar \a})
\label{basiscW}
\eq
forms for every $z,\bar z$ a basis for symplectic vectors.
More information about the expansion coefficients can be found in
appendix~\ref{s:basis}. This expansion will be used in the derivation of the
supersymmetric extension of the special K\"ahler constraints and of the field
equations.

\subsection{The constraint on the curvature}

Covariant derivatives involve the \Ka\ connection as in \eqn{defD},
and after choosing a real \Ka\ potential one may define a \Ka\
weight\footnote{$V$ and $U_\alpha$ have weight 1, while $Z_z$ has
weight 2, and for their complex conjugates respectively $-1$ and $-2$.}
$p$ for a symplectic section $W$, such that
\begin{equation}
{\cal D}_\alpha W= \left( \partial_\alpha +\ft p2
(\partial_\alpha K)\right) W\,,\qquad
{\cal D}_{\bar \alpha} W= \left( \partial_{\bar \alpha} -\ft p2
(\partial_{\bar \alpha} K)\right) W \,.  \label{covderp}
\end{equation}
If $W$ carries indices $\alpha$ or $\bar \alpha$ there is a further
metric connection, defined such that
${\cal D}_\alpha g_{\beta\bar \gamma}=0$. The curvature of the special \Ka\
manifold is then defined by
\bq
[\D_\a, \D_{\bar\b}] X_\g = - p g_{\a\bar\b} X_\g -
R_{\a\bar\b\g}{}^\d X_\d\,,
\eq
where $X_\a$ is a generic vector with K\"ahler weight $p$.
Applying this for $X_\alpha$ replaced by $U_\alpha$ and taking the
symplectic inner product with $\bar U_{\bar \delta}$ one finds
\bq
R_{\a\bar\b\g\bar\d}\equiv g_{\d\bar\d} R_{\a\bar\b\g}{}^\d =
- 2 g_{(\a|\bar\b|}
g_{\g)\bar\d} - \rmi \<  \D_\a U_\g , \D_{\bar\b} \bar U_{\bar\d}\>\,.
\label{curvatureSK}
\eq
If we introduce a symmetric tensor
\bq
C_{\a\b\g} = \<  U_\a, \D_\b U_\g\>\,,  \label{defC}
\eq
and expand the last term of \eqn{curvatureSK} in the basis ${\cal W}$
according to appendix~\ref{s:basis} we obtain the following two
cases:\\
\noindent
$1. \underline{{\rm The\, generic\, case:}}$
\begin{equation}
{\cal D}_\alpha U_\beta=\rmi C_{\alpha\beta\gamma}g^{\gamma\bar
\gamma}\bar U_{\bar \gamma}\ ,
\end{equation}
and the curvature is constrained to
\bq
R_{\a\bar\b\g\bar\d} \equiv g_{\b\bar\b}R^\b_{\a\g\bar\d} = - 2 g_{(\a|\bar\b|}
g_{\g)\bar\d} + C_{\a\g\e} g^{\e\bar\e} \bar C_{\bar\b\bar\d\bar\e}.
\eq
\noindent
$2. \underline{{\rm The\, special\, case:}}$\newline
In a similar way one finds that in this case
\bq
\mathcal{D}_z U_z = \rmi g'^{z\zb}(C_{zzz}\bar U_{\zb} - g_{z\zb}\D_z Z_z \bar V')\,.
\eq
The curvature becomes
\bq
R_{z\bar zz\bar z}= -2 g_{z\bar z}^2
 - g_{z\bar z}(\D_z Z_z)g^{\prime z\bar z}({\cal D}_{{\bar z}}\bar Z_{\bar z} )
+ C_{zzz} g^{\prime z\bar z}\bar C_{\bar z\bar z\bar z}   \,.
\eq
\subsection{An adapted basis and metric for the special case}

When $Z_z\neq 0$, one may diagonalize the matrix of symplectic
products between $V,\bar V,U_z$ and $\bar U_{\bar z}$ by
defining
\begin{equation}
U'_z=U_z +\rmi Z_z\bar V  \ ; \qquad \bar U'_{\bar z}=\bar U_{\bar z}
-\rmi \bar Z_{\bar z}V  \,.
\end{equation}
We then have symplectic products
\begin{eqnarray}
&&\< V,\bar V\>=\rmi\,, \qquad \< V,U'_z\>= \< V,\bar U'_{\bar z}\>=
\< \bar V,U'_z\>
= \< \bar V,\bar U'_{\bar z}\>=0\,,\nonumber\\
&& \<\bar U'_{\bar z}, U'_z\>=\rmi (g_{z\bar z}-Z_z\bar Z_{\bar z})= \rmi g'_{z\bar z}
\,.
\end{eqnarray}
In this way we find the Hermitian metric $g'_{z\bar z}$ which is invertible
because of \eqn{physicalcond}, but is not the second derivative of
the \Ka\ potential $K$, used to define the covariant derivatives in
\eqn{covderp}. With this definition, covariant
derivatives on the above equations lead to
\begin{equation}
 \< {\cal D}_zU'_z,\bar V \>= \< {\cal D}_zU'_z,V \>= 0 \,.\label{DU12}
\end{equation}
The defining expressions for $U_z$ and $Z_z$ imply
\begin{equation}
{\cal D}_z \bar U_{\bar z}=g_{z{\bar z}}\bar V\,,\qquad
{\cal D}_{\bar z}  U_z=g_{z{\bar z}} V\,,\qquad
{\cal D}_z  \bar Z_{\bar z}= {\cal D}_{\bar z} Z_z=0\,,
\end{equation}
which in the new basis give
\begin{equation}
{\cal D}_z \bar U'_{\bar z}=g'_{z{\bar z}}\bar V-\rmi\bar Z_{\bar z} U'_z\,,\qquad
{\cal D}_{\bar z}  U'_z=g'_{z{\bar z}} V+\rmi Z_z\bar U'_{\bar z}\,.   \label{DbUmod}
\end{equation}
When we define ${\cal D}'$ with metric connection
such that ${\cal D}'_z g'_{z\bar z}=0$, all the above relations
remain valid for ${\cal D}'$, as the non-zero connections are just
$\Gamma_{zz}^z$ and $\Gamma_{\bar z \bar z}^{\bar z}$. The new
definition now implies
\begin{equation}
\< \bar U'_{\bar z},{\cal D}'_z U'_z\> =0\,.
\end{equation}
The analogue of \eqn{defC} is then the definition
\begin{equation}
C'_{zzz}\equiv \< U'_z,{\cal D}'_z U'_z\>=C_{zzz}\,.
\end{equation}
This leads again to
\begin{equation}
{\cal D}'_z U'_z =\rmi C_{zzz} g'^{z\bar z} \bar U'_{\bar z}\,.
\end{equation}
We define then the curvature based on the metric $g'$ by
\begin{equation}
[\D'_z, \D'_{\bar z}] X_z = - p g_{z\bar z} X_z -
R'_{z\bar zz}{}^z X_z\,.
\end{equation}
Observe that the first term has $g$ and not $g'$ as this is the \Ka\
curvature. Calculating as before the curvature $R'$ by replacing
$X_z$ with $U'$, and an inner product with $\bar U'$, the last terms
in \eqn{DbUmod} lead to extra terms such that we find
\begin{equation}
R'_{z\bar zz\bar z}\equiv  R'_{z\bar zz}{}^z g'_{z\bar z}=
-2 g_{z\bar z} g'_{z\bar z} +C_{zzz}g'^{z\bar z}\bar C_{\bar z\bar z\bar z}\,.
\end{equation}
Rephrasing as much as possible in terms of the metric $g'_{z\zb}$,
we thus recover another geometry as for other special \Ka\ models.
There is an essential difference in the product of metrics in
\eqn{curvatureSK} and here. We tried to extend our analysis in the
basis ${\cal W'}=(V, U'_z, \bar V , \bar U'_{\bar z})$, but ran into
problems with the transformation rules because we want $V$ to be the
lowest component of a chiral multiplet. So, it is not possible to
get rid of $Z_z\neq 0$ by choosing another basis while keeping a
section of chiral multiplets. The model with $Z_z\neq 0$ is really
another model compared to those studied in the past.
\par
For some calculations below, it is also useful to introduce another
basis with symplectic vectors orthogonal to $U$. That is, we
introduce
\begin{eqnarray}
V'=V +\rmi Z_zg^{z\bar z}\bar U_{\bar z} &\,,\qquad&
\bar V'=\bar V -\rmi\bar Z_{\bar z}g^{z\bar z} U_z\,,      \nonumber\\
\< V',\bar V'\> = \rmi(1-g^{z{\bar z}}Z_z\bar Z_{\bar z}) &\,,
\qquad&\< V',U_z\>= \< V',\bar U_{\bar z}\>
=\< \bar V',U_z\> = \< \bar V',\bar U_{\bar z}\>=0\,,\nonumber\\
 \<\bar U_{\bar z}, U_z\>=\rmi g_{z\bar z}\,.
\end{eqnarray}
\subsection{Supersymmetric extension of special K\"ahler constraints}
\label{ss:susyKaconstr}

It is clear that the constraint (\ref{VVbar}) breaks the superconformal
symmetry. The constraints and their supersymmetric partners therefore play the
role of gauge conditions for some of the superconformal symmetries.
The residual symmetry should then still contain the symmetries of  Poincar\'e
supergravity.
In this subsection we will derive the supersymmetric partners of the constraints
\eqn{VVbar}, \eqn{DbarV} and \eqn{UU},  and compute the
decomposition rule for the
resulting supergravity, i.e.\ the rule which gives the remaining symmetry as
a linear combination of the original, superconformal, symmetry.

\subsubsection{Gauge choices and decomposition rule}

Before we go to these constraints, we {\it break the
special conformal symmetry} by imposing a constraint on $b_\mu$:
\bq\label{EC}
K\mbox{-gauge:}\quad b_\mu =0\, .
\eq
This does
not alter the number of degrees of freedom as $b_\mu$ is pure gauge in the
Weyl multiplet (cf table~\ref{dofbeforeConstr}).
\begin{table}[ht]
\begin{center}
\begin{tabular}{|c|c|l|}\hline
fields&d.o.f.&comments\\ \hline\hline
\multicolumn{3}{|c|}{The Weyl multiplet ($24+24$)}\\ \hline
$e_\mu^{\ a}$&5&16 - 4(translation.) - 6(Lorentz) - 1(dilatation)\\ \hline
$b_\mu$&0&4 - 4(special conformal.)\\ \hline
$A_\mu$&3&4 - 1($U(1)$)\\ \hline
${\cal V}_{\mu}{}^i{}_j$&9&12 - 3($SU(2)$)\\ \hline
$\psi_\mu^i$&16&32 - 8($Q$-supersymmetry) - 8($S$-supersymmetry)\\ \hline
$T^{ij}_{ab}$&6&complex antiself-dual \\ \hline
$\chi_i$&8&\\ \hline
$D$&1&real scalar\\ \hline\hline
\multicolumn{3}{|c|}{Symplectic section of chiral multiplets
($16(2n+2)+16(2n+2)$)}\\ \hline
$V$&2(2n+2)&\\ \hline
$\tilde \O_i$&8(2n+2)&\\ \hline
$\tilde Y_{ij}$&6(2n+2)&\\ \hline
$\tilde \F^-_{ab}$&6(2n+2)&\\ \hline
$\tilde \Lambda_i$ & 8(2n+2)&\\ \hline
$\tilde C$ & 2(2n+2)&\\ \hline
\end{tabular}
\caption{Degrees of freedom in the model before the constraints.
\label{dofbeforeConstr}}
\end{center}
\end{table}
\par
The decomposition rule for the special conformal symmetry is
\bq
\Lambda_K^a = -e^{\mu a} \left( \ft12 \be^i \phi_{\mu i} - \ft34 \be^i
\g_\mu \chi_i - \ft12 \bar \eta^i \psi_{\mu i} + \mbox{h.c.}\right)\,.
\label{LK}
\eq
Constraint (\ref{VVbar}) {\it breaks the dilatations}. Indeed, the superconformal
transformation of (\ref{VVbar}) gives
\bq
\<  \bar V, \be^i \tilde \O_i \>  - \<  V,\be_i\tilde
\O^i\> + 2\rmi \Lambda_D =0\,, \label{DecompD}
\eq
and the dilatations are now a combination of
other symmetries.
We choose as $S$-gauge
\bq
S\mbox{-gauge:}\quad\<  \bar V, \tilde \O_i\>  =0  \quad {\rm and} \quad
\<  V ,\tilde
\O^i\> =0\,.\label{Sgauge}
\eq
Remark that after this gauge choice the decomposition rule \eqn{DecompD}
simplifies to
\bq
\Lambda_D = 0\label{LD}\,,
\eq
such that we can forget about the original dilatations completely.
Demanding that the sections $V$ depend on $z^\a$ and $\bar z^{\bar
\a}$ in the way described in \eqn{defD}--\eqn{DbarV}, is a
{\it gauge choice for the chiral $U(1)$-transformations}. In fact,
consider the transformation of the first line of \eqn{transfo} using
these equations:
\bq
\d V = U_\a\d z^\a -\ft 12 (\der_\a K' \d z^\a
- \der_{\bar \a} K' \d \bar z^{\bar \a})V\,.
\eq
 An inner product with $\bar V$ gives (using (\ref{VVbar}) and its covariant
 derivative) a decomposition rule for the $U(1)$-transformations,
i.e.\
\bq
\Lambda_A= \Im(\der_\a K' \d z^\a)\,,
\label{LA}
\eq
where we have already used \eqn{kappa}.
\par
The decomposition rule for $\d_S(\eta_i)$ follows from the variation of the
$S$-gauge:
\brr
\eta_i &=& -\rmi \<  \bar V, \Dslash V\>  \e_i - \ft 12\rmi \<  \bar V,
\tilde Y_{ij}\>  \e^j \nonumber\\
& &- \ft 12\rmi \<  \bar V, \tilde {\cal F}_{ab}^-\>  \ve_{ij} \s^{ab} \e^j
- \rmi \<  \be_j \tilde \O^j, \tilde \O_i\> \,.
\label{SS}
\err
{}From now on, we only request that the constraints are invariant under the
resulting Poincar\'e
supersymmetry
\bq
\d(\e_i) = \d_Q(\e_i) + \d_S (\eta_i) + \d_A
(\Lambda_A) + \d_K(\Lambda_K)\,,
\eq
with $\Lambda_K$,  $\Lambda_A$ and $\eta_i$ defined in
\eqn{LK},  (\ref{LA})
and (\ref{SS}).
\par
Having the symplectic sections as functions of $z$ and $\bar z$, we
can consider the transformations of the bosonic constraints
\eqn{VVbar}--\eqn{DU=0}
and \eqn{UU}. The variation of the first one determined the breaking
of dilatations.  The constraints \eqn{DbarV} and \eqn{DU=0} are used to
determine the $z,{\bar z}$
dependences of $V$, $U$ and $K$ and their supersymmetry transformations are
thus trivial if we compute them in terms of $\delta z$ and $\delta\bar z$.
The constraint (\ref{UU}) is only non-trivial for $n>1$. Its
variation is
\begin{equation}
\d \<  U_\a, U_\b \>  = 2\<  {\cal D}_\g
U_{[\a},U_{\b]}\>  \d z^\g\,,
\end{equation}
which is $0$ due to the symmetry of \eqn{defC}.
This finishes the supersymmetry variations of the
bosonic special \Ka\ constraints.

\subsubsection{Physical fermions and fermionic constraints}

The first line of \eqn{transfo}, using \eqn{LD} and \eqn{LA}, is in terms of
$\delta z$:
\begin{equation}
\bar \epsilon^i \Omega_i=U_\alpha \delta z^\alpha\,.
\end{equation}
Therefore, the supersymmetry transformation of $z$ is chiral, and
we define $\l_i^\a$ as
\bq
\be^i \l_i^\a \equiv\d z^\a \,,
\eq
leading to
\bq
\tilde \O_i = U_\a \l_i^\a \,,
\label{Oml}
\eq
compatible with the $S$-gauge.
The relation (\ref{Oml}) can be inverted to
\bq
\l_i^\a = -\rmi g^{\a\bar \a}\< \bar U_{\bar \a} , \tilde \O_i\> \,.
\label{lambdainOmega}
\eq
That $\tilde \Omega_i$ has only components in the $U$ direction implies the constraints (the primes here and below are irrelevant for $n>1$ or $Z_z=0$)
\bq
\< V ,\tilde \O_i\>  =Z_z\lambda_i^\alpha \, \mbox{ or }\,
\< V' ,\tilde \O_i\>  = 0\,,\qquad
\<  U_\a ,\tilde \O_i\>  =0\,.
\label{UO}
\eq
The transformation rules for $z^\a$ and $\l_i^\a$ are\footnote{In the
 transformation laws below, there is still the $SU(2)$ transformation
 which is not gauge fixed and thus independent of the other
 transformations. We will not indicate these transformations explicitly, as they follow from
 the position of the $i$ indices.}
\brr
\d z^\a &=& \be^i \l_i^\a\,,\nonumber\\
\d \l_i^\a &=& - \Gamma^\a_{\b\g} \l_i^\b \d z^\g + \ft14 (\der_\b K \d
z^\b - \mbox{h.c.})\l^\a_i\nonumber\\
&&+\nslash z^\a \e_i - \ft 12\rmi g^{\a\bar \a} \<  \bar U_{\bar \a}, \tilde
Y_{ij}\>  \e^j - \ft 12\rmi g^{\a \bar \a} \<  \bar U_{\bar \a} ,\tilde {\cal
F}_{ab}^-\>  \ve_{ij} \s^{ab} \e^j\,,
\label{Psusy}
\err
where
\bq
\nabla_\mu z^\a = \der_\mu z^\a - \bar \psi^i_\mu \l_i^\a\,.
\eq

\subsubsection{Further variations of constraints in the generic case}

At the first fermionic level we have imposed the gauge
choice \eqn{Sgauge}, and found furthermore the constraints \eqn{UO},
leaving $n$ physical fermions as shown in \eqn{Oml} and \eqn{lambdainOmega}.
The variation of the $S$-gauge leads to the decomposition rule. Here
we will determine the further constraints on the $2(n+1)$ chiral multiplets
in the symplectic vector. We first perform this analysis for the
generic case where  $\<  V , U_\a\> = 0$, and treat the case $n=1$ separately
afterwards.
\par
The Poincar\'e transformations on (\ref{UO}) give
\brr
 \<  V, \tilde Y_{ij}\>   &=&0\,, \nonumber\\
\< U_\a , \tilde Y_{ij}\> & =& -C_{\a\b\g} \bar \l_i^\b \l_j^\g\,, \nonumber\\
 \<  V, \tilde \F_{ab}^-\>&=&0 \,,\nonumber\\
\<  U_\a ,\tilde {\cal F}^-_{ab}\>
   &=& -\ft12 C_{\a\b\g} \ve^{ij} (\bar\l_i^\b \s_{ab}\l_j^\g)\,.
\label{dUOF}
\err
To analyse the content of these equations, we make use of
lemma~B.1 of \cite{CRTVP}. This says that the
$2(n+1)\times (n+1)$ matrix $(V,U_\alpha )$ has rank $(n+1)$. Thus we
can solve (\ref{dUOF}) for half of the components of
$\tilde Y_{ij}$ and $\tilde \F_{ab}^-$.
\par
Straightforward variation of these two equations under Poincar\'e
supersymmetry yields a set of new constraints:
\brr
\<  V,\tilde\Lambda_i\>&=& - \ft16 C_{\a\b\g} \ve^{kl} (\bar \l_k^\a \s^{ab}
\l_l^\b) \s_{ab} \l_i^\g\,,\label{condL3}\\
\<  U_\a, \tilde \Lambda_i \> &=&
\ft 12\rmi C_{\a\b\g}  g^{\b\bar\b} \left( \< \bar U_{\bar \b} , \tilde Y_{ij} \>
\ve^{jk} \l_k^\g + \<  \bar
U_{\bar\b} , \s\cdot \tilde \F^-\>  \l_i^\g\right)\nonumber\\
& & + \ft16 {\cal D}_\a C_{\b\g\d} \cdot \ve^{kl}
(\bar \l_k^\b \s_{ab} \l^\g_l) \s^{ab} \l_i^\d\,. \label{condL1}
\err
Varying constraint (\ref{condL3}) yields
\brr
 \< V,\tilde C\>&=&  \ft 12\rmi \ve^{ik} \ve^{jl} C_{\a\b\g} g^{\a\bar\a}
\< \bar
U_{\bar\a} , \tilde Y_{ij}\>  \bar \l_k^\b \l_l^\g\nonumber\\
&&- \ft 12\rmi C_{\a\b\g} g^{\a\bar\a} \< \bar U_{\bar\a} , \tilde \F_{ab}^-\>
\ve^{kl} (\bar \l_k^\b \s^{ab} \l_l^\g)\nonumber\\
&&- \ft16 \D_\a C_{\b\g\d}\cdot \ve^{ij} (\bar \l_i^\a \s_{ab} \l_j^\b) \ve^{kl}
(\bar \l_k^\g \s^{ab} \l^\d_l)\,.\label{VC}
\err
The variation of (\ref{condL1}) gives
\brr\label{UC}
\< U_\a, \tilde C\>&=&  \ft14 C_{\a\b\g}  g^{\b\bar\b}
g^{\g\bar\g}\ve^{ik}\ve^{jl} \< \bar U_{\bar\b},
\tilde Y_{ij}\>  \< \bar U_{\bar \g}, \tilde Y_{kl}\> \nonumber\\
& & - \ft12  C_{\a\b\g} g^{\b\bar\b} g^{\g\bar\g}
\< \bar U_{\bar\b},
\tilde \F^-_{ab}\>  \< \bar U_{\bar \g}, \tilde \F^{-ab}\> \nonumber\\
& & - \ft 12\rmi \ve^{ik}\ve^{jl} [C_{\a\b\g}  \< \bar V, \tilde Y_{ij}\>
+ \D_\a C_{\b\g\d}\cdot g^{\d\bar\d} \< \bar U_{\bar \d}, \tilde
Y_{ij}\> ]\bar \l_k^\b \l_l^\g\nonumber\\
& & + \ft 12\rmi [ C_{\a\b\g}\< \bar V, \tilde \F_{ab}^-\>
+ \D_\a C_{\b\g\d}\cdot  g^{\d\bar\d} \< \bar U_{\bar \d}, \tilde \F_{ab}^-\> ]
\ve^{ij}(\bar \l_i^\b \s^{ab} \l_j^\g)\nonumber\\
& &-2\rmi C_{\a\b\g}  g^{\b\bar\b} \ve^{ij} \bar \l_i^\g \< \bar
U_{\bar \b}, \tilde \Lambda_j\>\nonumber\\
& & +\ft1{12} \D_\a \D_\b C_{\g\d\e} \cdot\ve^{ij}
(\bar \l_i^\b \s_{ab} \l_j^\g )\,\ve^{kl} (\bar \l_k^\d \s^{ab} \l_l^\e)\,.
\err
\par
These are all the possible \lq K\"ahler' constraints on the sections.
Let us review the degrees of freedom. Before imposing the
constraints, we have the degrees of freedom as in table~\ref{dofbeforeConstr}.
First of all there is the Weyl multiplet with $24+24$ degrees of freedom. The
gauge
invariances have been used to determine the counting. Indeed, the
dilation invariance can be seen as removing the
trace of the vierbein $e_\mu^\mu$ and $\gamma^\mu \psi_\mu^i$ is pure gauge
under special
supersymmetry. Similarly the vectors $A_\mu $ and ${\cal V}_\mu{}^j{}_
i$ lose a degree of freedom because of their gauge
transformations.
Secondly, we have the symplectic vectors of $2n+2$ chiral multiplets, which
altogether consist of $(2n+2) 16 + (2n+2) 16$ degrees of freedom.
\par
Then we have imposed the constraints \eqn{VVbar}--\eqn{DU=0}, \eqn{UU} and their
supersymmetry
partners. The new counting is in table~\ref{dofAfterConstr}. All the
symplectic sections are first reduced to $(n+1)$ rather than $(2n+2)$
degrees of freedom, as inner products with $V$ and with $U_\alpha $
are removed by the constraints. The symplectic vector $V$ is further
reduced to $n$ complex variables $z^\alpha $, by constraints which we have interpreted as
gauge choices
of  dilatations and chiral $U(1)$. These invariances have thus
disappeared, and in the upper part of the table we should thus
no longer subtract from degrees of freedom of the vierbein and of
$A_\mu $. Similarly, the constraint $\<  \bar V, \tilde \O_i\>=0$
removed a spinor doublet from the degrees of freedom of
$\tilde\Omega_i$, but this breaks the $S$--symmetry, and thus the
gravitino still has 24 degrees of freedom. As a result, the
superconformal invariance is reduced to super-\Poin. The
super-\Poin\ multiplet contains  the graviphoton, which
resides in $\< \bar V,\tilde \F^-_{ab}\> $. Similarly the other
internal products with $\bar V$ can be seen as auxiliary fields of
the $40+40$ off-shell super-\Poin\ multiplet. In other formulations
\cite{auxf,dWVHVP} they are expressed in terms of another
compensating multiplet. This compensating multiplet is then also used
to gauge fix the $SU(2)$ invariance which we have not broken here.
\begin{table}[ht]
\begin{center}
\begin{tabular}{|c|c|l|}\hline
fields&d.o.f.&comments\\ \hline\hline
\multicolumn{3}{|c|}{The Gravity Multiplet (40+40)}\\ \hline
$e_\mu^{\ a}$&6&16 - 4(translation) - 6(Lorentz)\\ \hline
$A_\mu$&4&gauge vector $\rightarrow$ vector\\ \hline
${\cal V}_{\mu}{}^i{}_j$&9&12 - 3($SU(2)$)\\ \hline
$\psi_\mu^i$&24&32 - 8($Q$-supersymmetry)\\ \hline
$T^{ij}_{ab}$&6&complex antiself-dual \\ \hline
$\chi_i$&8&\\ \hline
$D$&1&real scalar\\ \hline
$\< \bar V,\tilde Y_{ij}\> $&6&\\ \hline
$\< \bar V,\tilde \F^-_{ab}\> $&6&\\ \hline
$\< \bar V,\tilde \Lambda_i\> $ &8&\\ \hline
$\< \bar V,\tilde C\> $ &2&\\ \hline
\hline
\multicolumn{3}{|c|}{Symplectic section of constrained chiral multiplets
($16\,n+16\,n)$}\\ \hline
$z^\a$&2n&(d.o.f. of $V$)/2 - 2 (=trace vierbein + extra comp. of $A_\mu$)\\
\hline
$\lambda^\a_i$&8n&(d.o.f. of $\tilde \O_i$)/2 - 8 (=$\gamma^\mu \psi_\mu^i$)\\
\hline
$\< \bar U_{\bar\a},\tilde Y_{ij}\> $&6n&\\ \hline
$\< \bar U_{\bar\a},\tilde \F^-_{ab}\> $&6n&\\ \hline
$\< \bar U_{\bar\a} ,\tilde \Lambda_i\> $ &8n&\\ \hline
$\< \bar U_{\bar\a} ,\tilde C\> $ &2n&\\ \hline
\end{tabular}
\caption{Degrees of freedom in the model after the special \Ka\
constraints
\label{dofAfterConstr}}
\end{center}
\end{table}

\subsubsection{Further variations of constraints in the special case}

Now we continue the analysis of the supersymmetry transformations on
special K\"ahler constraints for
supergravity theories with $Z_z = \< V, U_z \> \neq 0$. This can only happen
for $n=1$, because that is the only case where this condition is not equivalent
with (\ref{UU}). Because $n=1$, $U_\a$ and $\D_\a$ can be replaced by $U_z$ and
$\D_z$.
\par
The computation of the special \Ka \ constraint goes along the same track as
for the generic case, but extra terms appear because of the weaker constraint.
The new contributions appear for the first time after the supersymmetry
variation of (\ref{UO}):
\brr
\< V', \tilde Y_{ij}\> &=& - \D_z Z_z\cdot \bar \l_i^z \l^z_j\,,\nonumber \\
\< U'_z , \tilde Y_{ij}\>  &=& -C_{zzz}  \bar \l_i^z \l_j^z\,, \nonumber \\
\< V', \tilde \F^-_{ab}\> &=& - \ft12 \D_z Z_z\cdot \ve^{ij}
(\bar \l_i^z \s_{ab}\l_j^z )\,,\nonumber\\
\<U'_z , \tilde {\cal F}^-_{ab}\>  &=&
 -\ft12 C_{zzz} \ve^{ij} (\bar \l_i^z \s_{ab}\l_j^z )\,.\label{dUOFn}
\err
Define a new vector $V''$ \footnote{Note that $g_{z\bar z} {\cal D}_z Z_z
= {\cal D}_{\bar z} C_{zzz} $ is not necessarily $0$ in this case.}
\begin{equation}
V'' \equiv V'- \frac{g^{z\bar z} {\cal D}_{\bar z}C_{zzz}\cdot U'_z}{C_{zzz}}
= V' - \frac{\D_z Z_z\cdot U'_z}{C_{zzz}}
\,.
\end{equation}
In terms of $V''$ the constraints will have the same form as before:
\begin{equation}
\<  V'' , \tilde Y_{ij}\> =
\<  V'' , \tilde \F_{ab}^- \>  = 0 \, .
\end{equation}
Then, one finds
\brr
 \<  V',\tilde\Lambda_i\>  &=&
\ft 12\rmi (\D_z Z_z) g^{z\bar z}(\< \bar U_{\bar z}, \tilde Y_{ij}\>  \ve^{jk} \l_k^z
+ \< \bar U_{\bar z}, \s \cdot \tilde \F^-\> \l_i^z)\nonumber\\
&& - \ft16 (C_{zzz} - \D_z\D_z Z_z)\,  \ve^{kl} \bar \l_k^z \s^{ab}
\l_l^z \s_{ab} \l_i^z\,,\label{condL3n} \\
\<  U'_z, \tilde \Lambda_i \>&=&
\ft 12\rmi C_{zzz}  g^{z\bar z}  (\<
\bar U_{\bar z} , \tilde Y_{ij} \>  \ve^{jk} \l_k^z + \<  \bar
U_{\bar z} , \s\cdot \tilde \F^-\>  \l_i^z)\nonumber\\
& & + \ft16 {\cal D}_z C_{zzz} \cdot \ve^{kl}
(\bar \l_k^z \s_{ab} \l^z_l) \s^{ab} \l_i^z\,,   \label{condL1n}
\err
where we have used that
\bq
 \<V,\D_z\D_zU_z\>  = -C_{zzz} + \D_z\D_z Z_z \,.
\eq
Note that these are the analogues of \eqn{condL1} and \eqn{condL3}.\\
Equation~\eqn{condL3n} can  be replaced by
\begin{equation}
\< V'', \tilde \Lambda_i\>  = \ft16  \,\ve^{kl} \bar
\l_k^z \s_{ab} \l_l^z \s^{ab} \l_i^z \frac{1}{C_{zzz}}( (-C_{zzz}
+ \D_z\D_z Z_z)\, C_{zzz}  - \D_z Z_z \cdot \D_z C_{zzz} )\,.
\end{equation}
Using the notation
\brr
{\cal O}_{zzzzz} &\equiv & g^{\prime z\bar z} \bar Z_{\bar z} \left[(-C_{zzz}
+ \D_z\D_z Z_z)\, C_{zzz}  - \D_z Z_z \cdot \D_z C_{zzz} \right] \, ,
\err
the variation of (\ref{condL1n}) now gives
\brr
\< U'_z, \tilde C\> &=&
\ft14 C_{zzz}  g^{z\bar z} g^{z\bar z}\ve^{ik}\ve^{jl}
\< \bar U_{\bar z},
\tilde Y_{ij}\>  \< \bar U_{\bar z}, \tilde Y_{kl}\> \nonumber\\
& & - \ft12 C_{zzz}  g^{z\bar z} g^{z\bar z}
\< \bar U_{\bar z},
\tilde \F^-_{ab}\>  \< \bar U_{\bar z}, \tilde \F^{-ab}\>\nonumber\\
& & - \ft 12\rmi \ve^{ik}\ve^{jl} ( C_{zzz}  \< \bar V, \tilde Y_{ij}\>
+ \D_z C_{zzz} \cdot g^{z\bar z} \< \bar U_{\bar z}, \tilde
Y_{ij}\> )
\bar \l_k^z \l_l^z\nonumber\\
& & + \ft 12\rmi ( C_{zzz}  \< \bar V, \tilde \F_{ab}^-\>
+ \D_z C_{zzz} \cdot g^{z\bar z} \< \bar U_{\bar z}, \tilde \F_{ab}^-\> )\,
\ve^{kl}(\bar \l_k^z \s^{ab} \l_l^z)\nonumber\\
& & -2\rmi C_{zzz}  g^{z\bar z} \ve^{ij} \bar \l_i^z \< \bar
U_{\bar z}, \tilde \Lambda_j\> \nonumber\\
& & +\ft1{12} ( \D_z \D_z C_{zzz}  + \ft12 {\cal
O}_{zzzzz})\, \ve^{ij} (\bar \l_i^z \s_{ab} \l_j^z) \ve^{kl} (\bar \l_k^z
\s^{ab} \l_l^z) \,.
\err
Straightforward variation of constraint (\ref{condL3n}) gives
\brr
\< V', \tilde C\> &=&
\ft14 \D_z Z_z \cdot g^{z\bar z} g^{z\bar z}\ve^{ik}\ve^{jl}
\< \bar U_{\bar z},
\tilde Y_{ij}\>  \< \bar U_{\bar z}, \tilde Y_{kl}\> \nonumber\\
& & - \ft12 \D_z Z_z  \cdot g^{z\bar z} g^{z\bar z}
\< \bar U_{\bar z},\tilde \F^-_{ab}\>  \< \bar U_{\bar z}, \tilde
\F^{-ab}\>\nonumber\\
& & + \ft 12\rmi \ve^{ik}\ve^{jl} \left((C_{zzz} - \D_z\D_z Z_z)
g^{z\bar z} \< \bar U_{\bar z}, \tilde Y_{ij}\> - \D_z Z_z \cdot
\< \bar V, \tilde Y_{ij}\> \right) \bar \l_k^z \l_l^z\nonumber\\
& & - \ft 12\rmi \left((C_{zzz} - \D_z\D_z Z_z)\,  g^{z\bar z} \< \bar U_{\bar z},
\tilde \F_{ab}^-\> - \D_z Z_z \cdot \< \bar V, \tilde \F_{ab}^-\>
 \right) \ve^{kl}(\bar \l_k^z \s^{ab} \l_l^z)\nonumber\\
& & - 2\rmi \D_z  Z_z \cdot g^{z\bar z}  \ve^{ij} \bar \l_i^z \< \bar U_{\bar z}, \tilde \Lambda_j\>
\nonumber\\
& & -\ft1{12} (2 \D_z C_{zzz} - \D_z \D_z \D_z Z_z)
\ve^{ij} (\bar \l_i^z \s_{ab} \l_j^z) \ve^{kl} (\bar \l_k^z \s^{ab} \l_l^z)\,.
\err
This can be rewritten in
\brr
\< V'', \tilde C\> & = & \frac{\rmi g^{z\zb}}{2C_{zzz}} \left[ (C_{zzz}
- \D_z\D_z Z_z)\, C_{zzz} + \D_z C_{zzz}\cdot\D_z Z_z\right] \nonumber \\
& &\times  \left[\ve^{ik}\ve^{jl}\<\bar U_{\zb}, \tilde Y_{ij}\>\bar \l^z_k\l^z_l
-\<\bar U_{\bar z}, \tilde \F_{ab}\>\ve^{kl} (\bar
\l^z_k\s^{ab}\l^z_l)\right]
\nonumber \\
& & -\ft{1}{12}\left[\left(2\D_z C_{zzz} - \D_z\D_z\D_z Z_z \right)
+ \frac{\D_z Z_z}{C_{zzz}}\left(\D_z \D_z C_{zzz}
+ \ft12 {\cal O}_{zzzzz}\right)\right] \nonumber \\
& & \times \ve^{ij} (\bar \l_i^z \s_{ab} \l_j^z) \ve^{kl} (\bar \l_k^z \s^{ab} \l_l^z)\,.
\err
\section{The generalized Bianchi identities combined with
the special \Ka\ constraints}\label{s:gBI}
In this section we start by imposing the reduction constraints on the chiral
multiplets. Because the constraints on the field strengths are Bianchi
identities, this linear multiplet of constraints is called the generalized
Bianchi identities. We combine
these constraints with the special K\"ahler constraints of section
\ref{s:kahler}. Together they give the field equations of $n$ vector
multiplets and expressions for the auxiliary fields $\chi_i$ and $D$.
We derive this for the generic case $\< V, U_\a \> =0$. We comment on
the supergravity equations of motion in this generic case. Finally,
we give the equations for the special case where $\< V, U_\a \> \neq
0$.

\subsection{The field equations for the generic case}\label{s:genBI}

To see what follows from equations~(\ref{bian1})--(\ref{bian4}), we take the symplectic
inner product of these equations with the basis ${\cal W}$.
The $4$ components of equation~(\ref{expn>1}) give $4$ equations for each
constraint.
\par
{}From the first identity we learn that the section $\tilde Y_{ij}$ is totally
constrained, as it should be because it is auxiliary. We have
\bq
\tilde Y_{ij} = -\rmi g^{\a\bar\a} \left[ C_{\a\b\g} \bar \l_i^\b \l_j^\g \bar
U_{\bar\a} -
\bar C_{\bar\a\bar\b\bar\g} \ve_{ik} \ve_{jl} \bar \l^{\bar\b k} \l^{\bar\g
l} U_\a \right ]\,.
\label{coY}
\eq
\par
It is more interesting to take a look at
(\ref{bian2}). Taking the symplectic inner product of this equation with
all components of the basis
${\cal W}$ (\ref{basiscW}), and using the special K\"ahler constraints of
section~\ref{ss:susyKaconstr} and \eqn{coY} gives:
\brr
\label{coL1}
\< \bar V, \tilde \Lambda_i\>  &=& 0\,,\nonumber\\
\< \bar U_{\bar \a} , \tilde \Lambda_i\>  &=& - \ve_{ij} \bar
C_{\bar\a\bar\b\bar\g} \nslash \bar z^{\bar \b}\cdot \l^{\bar\g j}\,,\nonumber\\
0 &=& \chi^i - \ft23 \s^{\mu\nu} ({\cal D}_\mu \psi^i_\nu - \ft18 \s\cdot
T^{ij} \gamma_\mu \psi_{\nu j}) - \ft12 g_{\a\bar\b} \dslash z^\a\cdot
\l^{i\bar\b} \nonumber\\
 && + \ft 12\rmi \g^\mu (\Qslash - \Aslash) \psi^i_\mu - \ft 14\rmi \ve^{ij} \g^\mu
\<  V,\s\cdot \tilde \F^+\>  \psi_{\mu\,j} + \ft 1{12}\rmi C_{\a\b\g} \ve^{ij}
\ve^{kl} (\bar \l^\a_k \s^{ab} \l_l^\b) \s_{ab} \l_j^\g\,,\nonumber\\
0 &=& \rmi g_{\a\bar\b} \left(\nslash \l^{i\bar\b} + \ft 12\rmi (\Qslash -
\Aslash)\lambda^{i\bar\b}\right) + \ft 12\rmi \ve^{ij} C_{\a\b\g} g^{\b\bar\b}\< \bar
U_{\bar\b}, \s \cdot \tilde \F^-\>  \lambda^\g_j \nonumber\\
&&+\ft 12\rmi C_{\a\b\g} \bar C_{\bar\a\bar\b\bar\g}g^{\g\bar\g}
(\bar\l^{i\bar\a}
\l^{j\bar\b}) \l_j^\b + \ft16 \D_\a C_{\b\g\d}\cdot \ve^{ij}\ve^{kl}
(\bar \l_k^\b
\s^{ab} \l_l^\g )\s_{ab} \l_j^\d\,,\label{bianchi2}
\err
where
\brr
\nabla_\mu \l_i^\a &=& \partial_\mu \l_i^\a -\ft12 \omega_\mu^{ab} \s_{ab}
\l_i^\a +\ft12 {\cal V}_{\mu\, i}^{\ \ \ j}\l_j^\a
+ \Gamma_{\b\g}^\a \partial_\mu z^\b \cdot\l_i^\g - \ft 12\rmi {\cal Q}_\mu
\l_i^\a\nonumber\\
&& - \nslash z^\a \cdot\psi_{\mu i} + \ft 12\rmi g^{\a\bar\a} \left[ \< \bar
U_{\bar\a}, \tilde Y_{ij}\>  + \ve_{ij} \< \bar U_{\bar \a}, \s \cdot \tilde
\F^-\>  \right] \psi_\mu^j\,,
\err
and
\bq
{\cal Q}_\mu = - \ft 12\rmi (\partial_\a K \cdot\partial_\mu z^\a - \mbox{h.c.})
\eq
is the K\"ahler 1-form.
\par
The first two equations in (\ref{bianchi2}) imply with (\ref{condL1}) and
(\ref{condL3}) that all components of $\tilde \Lambda_i$ are expressed
in terms of other fields, and thus they contain no independent degrees of
freedom.
The third expresses $\chi^i$ in terms of other fields.
In a superconformal calculation using a Lagrangian, this expression for
$\chi^i$
can be found from the equation of motion of the fermion of the compensating
vector multiplet.
The fourth equation is the
field equation for $n$ fermion doublets $\l^\b_i$.\vspace{5mm}
\par
We now proceed with the analysis of (\ref{bian3}). We first repeat
that (\ref{dUOF}) implies that there are $(n+1)$ independent
antisymmetric tensors in the symplectic vector ${\tilde \F}_{ab}$.
Apart from these there is another antiself-dual
tensor in the Weyl multiplet $T^{ij}_{ab}$.
A few definitions make (\ref{bian3}) more
transparent. First define the combination in brackets as
\bq \label{FTsec1}
\tilde {\hat F}_{ab} = \tilde {\cal F}_{ab} + \ft14 V T_{ab\, ij}\ve^{ij}
+ \ft14 \bar V T^{ij}_{ab} \ve_{ij}\,.
\eq
Then we take out covariantization terms:
\bq \label{FHsec1}
\tilde F_{ab} =\tilde {\hat F}_{ab} - 2 (\bar {\tilde \O^i} \g_{[a}
\psi_{b]}^j \ve_{ij}  + \bar V \bar \psi_a^i \psi^j_b \ve_{ij} +
\mbox{h.c.})\,.
\eq
This is chosen such that covariant derivatives in (\ref{bian3})
can be rewritten as ordinary derivatives, and the equation reduces to
\bq
\partial_\mu \ve^{\mu\nu\rho\s} {\tilde F}_{\rho\s} = 0\,.
\label{realBianchi}
\eq
Applying this on the $n+1$ independent components of $\tilde F_{\mu \nu }$,
implies that they can be expressed in terms of $n+1$ vectors. The
other $(n+1)$ equations of (\ref{realBianchi}) are the equivalent of field equations
for these vectors. Here, it is clear how our formulation keeps the symplectic covariance.
Only in the interpretation do we distinguish one half of the equations
as Bianchi identities and the others as field equations. These could
have been interchanged giving the `magnetic dual formulation'. Also
the fact of whether or not a prepotential exists is hidden here. The
difference is seen only when breaking the symplectic formulation in
finding an explicit solution of equations~(\ref{dUOF}).
If the $(n+1) \times (n+1)$-matrix, formed by the upper part of
$(V\,,U_\alpha )$
is invertible, then (\ref{dUOF}) expresses the $(n+1)$
lower components of $\F^-_{ab}$ in terms of the upper ones. This is
the case where there is a prepotential. When this matrix is not invertible\footnote{ As proven in \cite{CRTVP}, it is only the matrix
$\left(f_\a^I  \bar X^I\right)$ that is always invertible.}, then one
can still solve (\ref{dUOF}) for other $(n+1)$
components of $\tilde \F^-_{ab}$.
\par
We thus conclude that we have $n+1$ on-shell vectors and their field
equations also depend on the $6$ degrees of
freedom of the tensor $T^{ij}_{ab}$ of the Weyl multiplet.\vspace{5mm}
\par
Now let us have a look at \eqn{bian4}. It
involves the covariant Laplacian,
\brr
\Box \bar V &\equiv& \eta^{mn} D_m D_n \bar V\nonumber\\
            &=& e^{-1} \partial_\mu (e D^\mu \bar V)
                + (b^\mu - \rmi A^\mu) D_\mu \bar V + f_\mu^\mu \bar V
                + 2\bar \psi^{i[\mu} \g_\mu \psi_i^{\nu]}D_\nu\bar V\nonumber\\
            & & - \bar \psi^\mu_i D_\mu \tilde \O^i
                + \ft12 \bar \phi_i^\mu \g_\mu \tilde \O^i
                + \ft18 \bar \psi^i_\mu \g^\mu \s\cdot T_{ij} \tilde \O^j
                - \ft32 \bar \psi^i_\mu \g^\mu \chi_i \bar V\,.
\label{boxVbar}
\err
To derive this expression, we used a theorem on covariant derivatives in the
second reference of \cite{sctc}.
We can again take the symplectic inner product of (\ref{bian4}) with ${\cal
W}$:
\brr\label{eomsc}
\<  \bar V , \tilde C\>  &=& 0\,,\nonumber\\
\<  \bar U_{\bar\alpha}, \tilde C\>  &=& - 2\bar C_{\bar\a\bar\b\bar\g}
\nabla_\mu \bar z^{\bar\beta} \cdot\nabla^\mu \bar z^{\bar \gamma} +\ft 14
\bar C_{\bar\a\bar\b\bar\g}(\bar\l^{k\bar\b}\s\cdot T_{kl} \l^{l\bar\g})
\,,\nonumber\\
0 &=& -2e^{-1}\partial_\mu \left(e({\cal Q}^\mu - A^\mu) \right)
+ 2\rmi g_{\a\bar\a}\partial_\mu
z^\a\cdot \nabla^\mu{\bar z}^{\bar\a}
-2\rmi g_{\a\bar\a}\partial_\mu z^\a \cdot(\bar
\psi_i^\mu \l^{i\bar\a}) \nonumber\\
&&-2\rmi({\cal Q}^\mu - A^\mu)({\cal Q}_\mu - A_\mu)
+2\rmi f_\mu^\mu -4{\bar \psi}^{i[\mu}\g_\mu\psi_i^{\nu]}({\cal Q}_\nu - A_\nu)
\nonumber\\
&&+{\bar \psi}_i^\mu \< V,\s\cdot \tilde \F^+ \>  \ve^{ij}\psi_{\mu j}
+2\rmi{\bar \psi}_i^\mu \phi^i_\mu
-2 \bar \psi_i^\mu(\Qslash - \Aslash)\psi^i_\mu \nonumber \\
&&-3\rmi{\bar\psi}^i_\mu \g^\mu \chi_i
+\ft 14 \< V, \tilde\F^+_{\mu\nu}\>  T^{\mu\nu}_{ij}\ve^{ij}
-\ft 12\rmi C_{\a\b\g}g^{\a\bar\a}
\bar C_{\bar\a\bar\b\bar\g}(\bar\l^{\bar\b k}\l^{\bar\g
l})(\bar\l^\b_k\l^\g_l)\nonumber \\
&&-\ft 12\rmi C_{\a\b\g}g^{\a\bar\a} \< {\bar U}_{\bar\a},
\tilde \F^-_{ab}\>  \ve^{kl}(\bar\l^\b_k \s^{ab} \l_l^\g)
-\ft 16 {\cal D}_\a C_{\b\g\d}\cdot\ve^{ij}(\bar\l^\a_i\s_{ab}\l_j^\b)\,
\ve^{kl}
(\bar\l^\g_k\s^{ab}\l_l^\d)\,,\nonumber\\
0 &=& -2\rmi e^{-1}g_{\a\bar\a}\partial_\mu (e \nabla^\mu \bar z^{\bar\a})
+4({\cal Q}_\mu - A_\mu)g_{\a\bar\a}\nabla^\mu\bar z^{\bar\a}
+g_{\a\bar\a}({\cal Q}_\mu - A_\mu)(\bar\psi_i^\mu\l^{i\bar\a})
\nonumber \\
&&-2\rmi g_{\alpha \bar \alpha }\Gamma_{\bar\b\bar\gamma }^{\bar\alpha }
\partial_\mu \bar z^{\bar\b}
\cdot\nabla^\mu{\bar z}^{\bar\gamma }
-4\rmi g_{\a\bar\a}{\bar \psi}^{i[\mu} \g_\mu \psi_i^{\nu]}\nabla_\nu\bar
z^{\bar\a} +2\rmi g_{\a\bar\a}\bar \psi_i^\mu\nabla_\mu \l^{i\bar\a}\nonumber \\
&&-\rmi g_{\a\bar\a}\bar\phi_i^\mu \g_\mu \l^{i\bar\a}
+2C_{\a\b\g}\ve^{ik}\ve^{jl}(\bar\l^\b_k\l^\g_l)(\bar \psi^\mu_i\psi_{j\mu})
+2\bar\psi^\mu_i \<  U_\a, \s \cdot \tilde\F^+\ve^{ij}\> \psi_{\mu j}\nonumber\\
&&-\ft 14\rmi g_{\a\bar\a}\bar\psi^i_\mu\g^\mu\s\cdot T_{ij}\l^{j\bar\a}
-6\rmi g_{\a\bar\a}\bar \chi_i\l^{i\bar\a}\nonumber \\
&&+\ft 14 \< U_\a, \tilde \F^+_{\mu\nu}\>  T^{\mu\nu}_{ij}\ve^{ij}
+\ft 14 C_{\a\b\g}g^{\b\bar\b}g^{\g\bar\g}\ve_{ik}
\ve_{jl}\bar C_{\bar\b\bar\d\bar\ve}
(\bar\l^{i\bar\d}\l^{j\bar\ve})\bar C_{\bar\g\bar\a\bar\zeta}
(\bar\l^{k\bar\a}\l^{l\bar\zeta}) \nonumber \\
&&-\ft 12 C_{\a\b\g}g^{\b\bar\b}g^{\g\bar\g}\< \bar
U_{\bar\b}, \tilde \F^-_{ab}\> \< \bar U_{\bar\g}, \tilde \F^{-ab}\>
-\ft 12\rmi {\cal D}_\a C_{\b\g\d}\cdot g^{\d\bar\d}\bar C_{\bar\d\bar\b\bar\g}
(\bar\l^{k\bar\b}\l^{l\bar\g})(\bar\l_k^\b\l^\g_l) \nonumber \\
&&+\ft 12\rmi \left[ C_{\a\b\g}\< \bar V, \tilde \F^-_{ab}\>
+{\cal D}_\a C_{\b\g\d}\cdot g^{\d\bar\d}
\< \bar U_{\bar\d}, \tilde \F^-_{ab}\> \right]
\ve^{ij}(\bar\l^\b_i\s^{ab}\l_j^\g)  \\
&&+\ft 1{12} {\cal D}_\a{\cal D}_\b C_{\g\d\ve}
\cdot\ve^{ij}(\bar\l^\b_i\s_{ab}\l_j^\g)
\ve^{kl} (\bar\l^\d_k\s^{ab}\l_l^\ve)
-2\rmi C_{\a\b\g}g^{\b\bar\b}\bar C_{\bar\b\bar\g\bar\d}\bar\l^\g_i\,\nslash\bar
z^{\bar\g}\cdot\l^{i\bar\d}\,.\nonumber
\err
The first two equations, together with (\ref{UC}) and (\ref{VC})
(which can be simplified using the second equations in (\ref{bianchi2}))
determine that $\tilde C$ is completely determined in terms of other
fields.
The real and  imaginary part of the third equation have both to be $0$.
The real part constrains the divergence of $\mathcal{Q}_\mu - A_\mu$, and
the imaginary part gives an expression for the $D$-field of the Weyl multiplet.
($D$ is hidden in the $f^\mu_\mu$-term.)
The fourth equation of the expansion in terms of ${\cal W}$ gives the field
equations for $n$ complex scalars.
So we find the same structure in the equations as for the fermions: $n+1$
equations express $\tilde C$ in terms of other fields, while the $n+1$ other
equations give the field equations for $n$ complex scalars $z^\a$, an
expression for $D$ and a constraint for $({\cal Q}_\mu - A_\mu)$. The degrees of freedom are described in table~\ref{dofAfterConstrBianchi}.
\begin{table}[ht]
\begin{center}
\begin{tabular}{|c|c|l|}\hline
fields&d.o.f.&comments\\ \hline\hline
\multicolumn{3}{|c|}{The Gravity Multiplet}\\ \hline
$e_\mu^{\ a}$&6&16 - 4(translation) - 6(Lorentz)\\ \hline
$A_\mu$&3&$\partial ^\mu A_\mu$ constrained\\ \hline
${\cal V}_\mu{}^j{}_i$&9&12 - 3($SU(2)$)\\ \hline
$\psi_\mu^i$&24&32 - 8($Q$-supersymmetry)\\ \hline
$T^{ij}_{ab}$&6&$n$ on-shell and $2$ off-shell vectors \\ \hline
$\chi_i$&0& expressed in terms of other fields\\ \hline
$D$&0&expressed in terms of other fields\\ \hline\hline
\end{tabular}
\caption{Off-shell degrees of freedom after the special \Ka\
constraints and the generalized Bianchi identities: $24 + 24$ d.o.f..
All other variables are expressed in terms of these fields or have a
field equation (for $n$ complex scalars, $n$ doublet spinors and
$n+1$ vectors).
\label{dofAfterConstrBianchi}}
\end{center}\end{table}

\subsection{Comments on the supergravity equations}

Using the results of the previous section, we comment on the appearance of
equations of motion for the remaining $24+24$ components of
table~\ref{dofAfterConstrBianchi} from one symplectic invariant constraint
\bq
\< V , \tilde {\cal F}^+_{ab} \>  \approx 0\label{FET}\,,
\eq
which gives rise to a $24+24$ `current' multiplet.
The $\approx$-sign is used to denote that we only expose the linear terms.
This already shows the essential features of this symplectic covariant
formulation. With the linearized approximation we mean that we
keep terms with an arbitrary power of undifferentiated scalar
fields or metric, but only linear in other fields. In a full
treatment of $N=2$ supergravity couplings the right-hand side
 of (\ref{FET}) would, for example,
contain an additional coupling to hypermultiplets.
\par
To discuss the supersymmetry partners of (\ref{FET}),
we derive a new $N=2$ multiplet with 24 + 24 components.
The multiplet starting with the symplectic expression $\<V , \tilde
{\cal F}^+_{ab} \> $ is a supergravity realization of this multiplet.
As shown below the supermultiplet of constraints derived from (\ref{FET})
is only equivalent to the supergravity equations of motion, up to
integration `constants'. These $8+8$ remaining unknowns can be determined when
one of the three possibilities of a second compensating multiplet is
introduced as in \cite{auxf,dWVHVP, dWLVP}. In our approach this is the
place where the second compensating multiplet, which is also needed for
consistency in the Lagrangian formulation, comes into play.

\subsubsection{A restricted chiral self-dual tensor multiplet}
The supermultiplet structure of the `current' multiplet from (\ref{FET}) is
that of  a chiral self-dual tensor multiplet,
\brr
W_{ab}^+ &=& A_{ab}^+ + \bar \t^i \psi_{abi} + \ft14 \bar \t^i \t^j B_{abij} + \ft14 \ve_{ij} \bar \t^i \s_{cd} \t^j F_{ab}{}^{cd}
\nonumber\\ &&
+ \ft16 \ve_{ij} (\bar \t^i \s_{cd} \t^j) \bar \t^k \s^{cd} \chi_{abk}
+ \ft1{48} (\ve_{ij} \bar \t^i \s_{cd} \t^j)^2  C^+_{ab}\,. \label{chasdsup}
\err
It has the following field content.
$A_{ab}^+$ is a self-dual complex tensor with 6 degrees of freedom. $\psi_{abi}$ has 24
left-handed fermionic components. The tensor $B_{abij}$ has 18 components.
The tensor $F_{ab}{}^{cd}$ is self-dual in its first and
antiself-dual in its second pair of indices, leading to 9 complex
components. It also satisfies the following properties:
\begin{eqnarray}
F_{ab,cd}+F_{cd,ab}
&=& \ft12 \varepsilon _{ab}{}^{ef}\left( F_{ef,cd} -
F_{cd,ef}\right)\,,
\nonumber\\
F_{ab,cd}-F_{cd,ab}
& = & \varepsilon _{abe[d}\left( F^e{}_{c]} +F_{c]}{}^e\right)\,,
\nonumber\\
F_{ab,cd}
&=& \delta_{a[c}F_{d]b} -\delta_{b[c} F_{d]a} -
\ve_{abe[c}F^e{}_{d]}\,,\nonumber\\
F_{[ab]} & = &  0 \nonumber\,,\\
  F^a{}_{a} & = & 0\label{Fprop} \,,
\end{eqnarray}
where
\begin{equation}
F_a{}^c = F_{ab}{}^{cd} \delta^b_d\,.
\end{equation}
A general component of this self-dual--antiself-dual tensor
$F_{ab}{}^{cd}$ can thus be written in terms of the traceless
symmetric part $F_{(ab)}$ with 9 components. The fermion $\chi_{abi}$
has again 24 left-handed components and $C_{ab}^+$ has 6. So, this is
a chiral multiplet with $48 + 48$ components.
\par
The transformation rules of this multiplet are the same as
for a chiral multiplet with a complex scalar as lowest component, but
with the components replaced straightforwardly:
\brr
\d A^+_{ab} & = & \be^i \psi_{abi}\,, \nonumber \\
\d \psi_{abi} &=& \dslash A_{ab}^+ \epsilon_i + \ft12 B_{abij}\epsilon^j  + \ft12
                   \s_{cd}  F_{ab}{}^{cd} \ve_{ij} \epsilon ^j\,, \nonumber\\
\d B_{abij} &=& 2 \be_{(i} \dslash \psi_{abj)}
- 2\be^k \chi_{ab(i} \ve_{j)k}\,,\nonumber\\
\d F_{ab}{}^{cd}&=&\ve^{ij} \be_i \dslash \s^{cd} \psi_{abj}
+ \be^i \s^{cd} \chi_{abi}\,, \nonumber\\
\d \chi_{abi} &=&-\ft12 \s_{cd} F_{ab}^{cd} \stackrel{\leftarrow} {\dslash}
           \epsilon_i - \ft 12 \dslash B_{abij} \ve^{jk} \epsilon_k + \ft12 C_{ab}^+
                       \ve_{ij}\epsilon^j\,,  \nonumber\\
\d C_{ab}^+ &=& - 2 \ve^{ij} \be_i \dslash \chi_{abj}\,.\label{susy}
\err
Since we have broken superconformal symmetry to super-Poincar\' e and $SU(2)$,
we only need a super-Poincar\' e version of this multiplet.
Note that it cannot be extended to a superconformal one. The
commutator of a supersymmetry and a special supersymmetry has to give
a Lorentz transformation that can never be realized because of the
duality and chirality properties of the spinors. For this reason,
it is only possible to construct an antiself-dual chiral tensor multiplet,
realizing the superconformal algebra, as given in \cite{BDRDW}.
\par
To study the field equations of the fields of table~\ref{dofAfterConstrBianchi}, we need a multiplet with $24+24$ components.
A suitable multiplet of constraints is:
\brr
0 & = & \partial^a(B_{abij} + \ve_{ik}\ve_{jl}{\bar B}_{ab}^{kl})\,, \label{conB}\\
0 & = & \partial^a( \chi_{ab}^i -\ve^{ij}\dslash \psi_{abj})\,, \label{conchi}\\
0 & = & \partial^a( C^-_{ab} - \Box A^+_{ab})\,,\label{conC}\\
0 & = & \partial^a\partial_c(F_{ab}{}^{cd} + {\bar F}_{ab}{}^{cd})\,. \label{conF}
\err
These are the analogues of the constraints (5.4) in \cite{BDRDW}.
This set contains  $(9+6+9) + 24$ equations.
The constraint for $F_{ab}{}^{cd}$ splits up in a part symmetric in $(bd)$
(6 independent equations) and an antisymmetric part in $[bd]$ (3 independent
equations), which correspond to the real and imaginary part of $F_{ac}$:
\brr\label{conF'}
0 & = & -\partial^c\left(\partial_{(b}(F_{d)c} +{\bar F}_{d)c})\right)+\ft 12 \delta_{bd}\partial^a\partial^c(F_{ac}+{\bar F}_{ac})+\ft12 \Box(F_{bd}+{\bar F}_{bd})\nonumber \\
& & +\ft 12\ve_{bdae}\partial^a\partial^c(F^e{}_c - {\bar
F}^e{}_c)\,.
\err
As far as we know, this reduced multiplet is a new representation of the rigid $N=2$ algebra.

An explicit supergravity realization of this reduced multiplet
is given by
\begin{eqnarray}
A^+_{ab} &=& \langle V, {\tilde {\cal F}}_{ab}^+\rangle\,,\nonumber\\
\psi_{abi} &\approx& -\rmi\ve_{ij}\g^\rho\s_{ab}\phi_\rho^j\,,\nonumber\\
B_{abik} & \approx & 2\rmi\ve_{ij}R_{SU(2)}{}_{ab}^+{}^j{}_k\,,\nonumber \\
F_{ab}{}^{cd} & \approx & 2 \delta^{[c}_{[a} \left(\partial_{b]}({\cal Q}^{d]}-A^{d]})
+ (\partial^{d]}({\cal Q}_{b]}-A_{b]})) -2\rmi{\cal R}_{b]}{}^{d]}
+\ft 12\rmi\delta_{b]}^{d]}{\cal R}\right)\nonumber \\
& & -\ve^{cd}{}_{ef}\delta^{[e}_{[a} \left(\partial_{b]}({\cal Q}^{f]}-A^{f]})
+ (\partial^{f]}({\cal Q}_{b]}-A_{b]})) -2\rmi {\cal R}_{b]}{}^{f]}
+\ft 12\rmi\delta_{b]}^{f]}{\cal R}\right)\,.
\label{currmult}
\end{eqnarray}
In deriving this multiplet we used the constraints of
sections~\ref{s:kahler} and \ref{s:gBI}.
The expression for $B_{abij}$ satisfies constraint (\ref{conB}), which
is a Bianchi identity that expresses the existence of $SU(2)$-vectors.
The expression for $F_{ab}{}^{cd}$ fulfils
(\ref{Fprop}). It also satisfies (\ref{conF})
when the third equation of \eqn{eomsc} for $({\cal Q}_\mu - A_\mu)$ is used.
Therefore, the multiplet derived from $\langle V,{\tilde{\cal F}}_{ab}^+\rangle$
has 24 + 24 components.

\subsubsection{Some comments on the multiplet of equations from
$\langle V,{\tilde{\cal F}}_{ab}^+\rangle \approx 0$}

Putting the `current' multiplet (\ref{currmult}) to zero, will give
rise to some supergravity field equations.
These are 24 + 24 equations for the 24 + 24 remaining degrees of freedom of
table~\ref{dofAfterConstrBianchi}. The counting in this table
subtracts the gauge degrees of freedom.  The multiplet here is a multiplet
of curvatures and the counting is equivalent if we take into account the
Bianchi identities.
\par
However, our equations are not equivalent to the complete supergravity
equations of motion. They differ modulo `integration constants'. These can
be determined when a second compensating multiplet is coupled
\cite{dWVHVP,auxf}. Since this step is independent
of the symplectic formulation of the coupling of vector multiplets to
supergravity, we do not treat it here.
\par
Let us give a brief discussion of the content of the equations following
from \eqn{FET}. Equation \eqn{FET} reduces 6 degrees of freedom.
It expresses the `graviphoton' field strength $T_{abij}$ as a combination of the
$n+1$ on-shell vectors obtained above. It is the symplectic
expression for the algebraic equation of motion that one finds
in the Lagrangian approach, (4.11) in \cite{dWLVP}.

Using \eqn{bianchi2} in (\ref{dependent}) with
\bq
R^{\mu i} \equiv e^{-1}\ve^{\mu\nu\rho\sigma}\g_5\g_\nu
\left({\cal D}_\rho\psi_\s^i -\ft 18 \s\cdot T^{ij}\g_\rho\psi_{\s j}\right)
\eq
in the second component of the current multiplet gives that
\bq\label{phi0}
\phi_\rho^i = R_\rho^i -\ft 14 \g_\rho\g\cdot R_\rho^i \approx 0\,,
\label{FEpsi}
\eq
the traceless part of the field equation of the gravitini.
Therefore, this equation cannot determine the trace-part $\gamma\cdot R^i$.
However, combining \eqn{FEpsi} with the Bianchi identity for the
gravitino field strength $\partial^\mu R_\mu^i \approx 0$, yields
\begin{equation}
\dslash \gamma\cdot R^i\approx 0\,,
\end{equation}
which determines $\gamma \cdot R^i$ in terms of 8 `integration constants'.

The $B_{abij}$ component yields
\begin{equation}
R_{SU(2)}{}_{ab}{}^i{}_j \approx 0\,.
\end{equation}
Together with the Bianchi identity for the $SU(2)$ curvature it states that
the gauge fields ${\cal V}_\mu{}^i{}_j$ are pure gauge, i.e.\
\begin{equation}
{\cal V}_\mu{}^i{}_j = (\varphi^{-1} \partial_\mu \varphi)^i{}_j\,,
\end{equation}
where $\varphi$ is a group element of $SU(2)$. The three local parameters defining
$\varphi$ are left undetermined.

$F_{ab}{}^{cd}$ has its components in the traceless part of $F_{(ac)}$.
{}From $F_{ab}{}^{cd} \approx 0$ follows
\bq
F_{ac} = 2\partial_{(a}({\cal Q}_{c)}-A_{c)}) - 2\rmi{\cal R}_{ac} +\ft 12\rmi
g_{ac}{\cal R}\approx 0\,.
\eq
The imaginary part is the traceless part of the Einstein equation. Again we
cannot determine the scalar curvature ${\R}$ from this equation. However,
combining this equation with the Bianchi identity for the Einstein tensor
\bq
\partial^a({\cal R}_{ab} -\ft 12 g_{ab}{\cal R}) = 0\,,
\eq
gives
\begin{equation}
\partial^a {\R} \approx 0
\end{equation}
and again $\R$ is determined up to a constant.
The real part of the $F$-component gives that
$A_\mu \approx {\cal Q}_\mu $ up to a constant vector.
Also in the Lagrangian approach \cite{dWLVP}, one finds
\bq
A_\mu \approx {\cal Q}_\mu \,.
\eq

The additional $8+8$ remaining unknowns can be determined through the field
equations of a second compensating multiplet. This concludes the short
discussion of the supergravity equations of motion.

\subsection{The field equations for the special case}

In this subsection, the expressions of section \ref{s:genBI} are generalized
for the case $n=1$ where further $Z_z = \< V, U_z \> \neq 0$. This is the
case that was excluded by the former definitions and where our less
restrictive definitions becomes
relevant. The equations are found by expanding the constraints in terms of the
basis of symplectic vectors using the methods mentioned at the end of the
appendix.\vspace{5mm}
\par
The section $\tilde Y_{ij}$ remains totally constrained:
\brr
\tilde Y_{ij} & = & g'^{z\bar z}\Big( -\rmi\ve_{ik}\ve_{jl}
 g_{z\zb}\D_{\zb} \bar Z_{\bar z}\cdot \bar\l^{k\zb} \l^{l\zb}V
 +\rmi g_{z\zb}\D_z Z_z\cdot\bar\l^z_i\l^z_j\bar V \nonumber\\
 && \hspace{1cm}+\rmi\ve_{ik}\ve_{jl} \bar C_{\bar z\bar z\bar z} \bar\l^{k{\zb}}\l^{l{\zb}}  U_z
 -\rmi C_{zzz}\bar\l^z_i\l^z_j\bar U_{\zb}\Big)\,.
\err
This equation reduces to the former equation~(\ref{coY}) when $\< V, U_z \> =0$. \vspace{5mm}
\par
The equations that can be derived from the constraints for the fermions are
the following ones:
\brr
\<\bar V',\tilde \Lambda _i\>&=&
- \D_{\zb}\bar Z_{\bar z}\cdot \ve_{ij} \nslash \bar z\cdot
\l^{j\zb}\,,
\nonumber \\
\<\bar U'_{\zb},\Lambda_i\>&=& - \ve_{ij} \bar C_{\bar z\bar z\bar z}
\nslash {\zb} \cdot\l^{j{\zb}}+ \ft12\gamma ^\mu\ve_{ij}
\left( \<\bar U'_{\zb}, \tilde Y^{jk}+\sigma \cdot\tilde \F^+\ve^{jk}\> \right)
\psi_{\mu k} \,,
\nonumber \\
0 & = & \chi^i - \ft23 \s^{\mu\nu} ({\cal D}_\mu \psi^i_\nu - \ft18 \s\cdot
T^{ij} \gamma_\mu \psi_{\nu j}) - \ft12 \left(g_{z\zb} \dslash z \cdot
\l^{i\zb} -i \g^\mu (\Qslash - \Aslash)\right) \psi^i_\mu \nonumber\\
& &- g'^{z\bar z}\left[ \ft12 Z_z \bar C_{\bar z\bar z\bar z}\dslash\bar z\cdot\l^{i\zb}
+ \ft 14\rmi \g^\mu (g_{z\zb}\< V',\tilde Y^{ij} + \s\cdot \tilde \F^+\ve^{ij}\>
\psi_{\mu\,j}\right. \nonumber \\
&&- \ft 14 \ve^{ij} \D_z Z_z\cdot (\<\bar U_{\zb},\tilde Y_{jk}\>\ve^{kl}\l^z_l
+\<\bar U_{\zb},\s \cdot\tilde \F^-\>\l^z_j) \nonumber \\
&&\left. - \ft 1{12}\rmi \ve^{ij} g_{z\zb} (C_{zzz} - \D_z\D_z Z_z)\,
\ve^{kl}(\bar \l^z_k\s_{ab}\l^z_l)\s^{ab}\l_j^z  \right] \,,\nonumber\\
0 & = & \rmi g'_{z\bar z}\left(\nslash \l^{i\zb} + \ft{1}{2}\rmi \left(\Qslash -
\Aslash\right)\lambda^{i\zb}\right)\nonumber\\ &&
+ \ft{1}{2}\rmi\ve^{ij} C_{zzz} g^{z\zb}\left(\< \bar
U_{\zb},\tilde Y_{jk}\>\ve^{kl}\l^z_l +\<\bar U_{\zb}, \s \cdot \tilde \F^-\>
\lambda^z_j\right) \nonumber\\
&& + \ft16 \D_z C_{zzz}\cdot \ve^{ij}\ve^{kl} (\bar \l_k^z
\s^{ab} \l_l^z) \s_{ab} \l_j^z -\rmi Z_z \D_{\zb}\bar Z_{\bar z}\cdot\nslash
\zb\cdot \l^{i\zb}\,.
\err
Also here, the equations reduce to those that we have found for the generic
case where $\<V,U_z\> = 0$. The same comments as in section \ref{s:genBI} are
valid here.\vspace{5mm}
\par
Repeating the analysis for the equations of the vectors, it appears that there
is no information used about $Z_z$. This means that the analysis of the equations
for the vectors of section \ref{s:genBI} remains valid. This is no surprise
because the
equations for the vectors are a symplectic section of equations. All the other
equations are singlets for the symplectic group and can therefore be written
as symplectic invariant equations.\vspace{5mm}
\par
Also the last constraint can be decomposed with respect to the symplectic basis.
Then the equations become:
\brr
 \< \bar V' , \tilde C\> &=&-2 \bar\psi_i^\mu\<\bar V,\tilde Y^{ij}
+\s\cdot\tilde\F^+\ve^{ij}\>\psi_{\mu j} -2 \partial_\mu\zb \cdot\D_{\zb}
\bar Z_{\bar z}\cdot \left(\nabla^\mu\zb-\bar\psi^\mu_i\l^{i\zb}\right)
 \nonumber \\
&& +\ft 14 \D_{\zb}\bar Z_{\bar z}\bar\l^{i\zb}\s
\cdot T_{ij}\l^{j\zb}\,,\nonumber\\
 \< \bar U'_{\zb}, \tilde C\>&=& - 2\bar C_{\bar z\bar z\bar z}
\nabla_\mu \zb \cdot \nabla^\mu \zb +\ft 14\bar C_{\bar z\bar z\bar z}
(\bar\l^{k\zb}\s\cdot T_{kl} \l^{l\zb})
\,,\nonumber\\
0 &=& g^{z\zb}g'_{z\bar z}
\Big( 2e^{-1}\partial_\mu \left(e({\cal Q}^\mu - A^\mu)\right)
+2\rmi({\cal Q}^\mu - A^\mu)({\cal Q}_\mu - A_\mu) -2\rmi f_\mu^\mu
+3\rmi{\bar\psi}^i_\mu \g^\mu \chi_i \nonumber \\
&&- 2\rmi g_{z\zb}\partial_\mu z
\cdot(\nabla^\mu{\bar z} -\bar \psi_i^\mu \l^{i\zb})\nonumber\\ &&
+4{\bar \psi}^{i[\mu}\g_\mu\psi_i^{\nu]}({\cal Q}_\nu - A_\nu)
+2 \bar \psi_i^\mu(\Qslash - \Aslash)\psi^i_\mu
-2\rmi{\bar \psi}_i^\mu \phi^i_\mu \Big)
\nonumber\\
&& +\bar\psi^\mu_i\psi_{\mu j} \ve^{ik}\ve^{jl}\D_z Z_z\cdot\bar\l^z_k\l^z_l
- {\bar \psi}_i^\mu \< V',\s\cdot \tilde \F^+ \>  \ve^{ij} \psi_{\mu j}
-\ft 14 g^{z\zb} g_{z\zb}\< V', \tilde\F^+_{\mu\nu}\> T^{\mu\nu}_{ij}\ve^{ij}
\nonumber \\
&& +2\rmi g^{z\zb}Z_z
\left(\rmi\D_{\zb}\bar Z_{\bar z}\cdot\partial_\mu\bar z\cdot
\left({\cal Q}_\mu - A_\mu\right) +\bar\psi^\mu_i\bar C_{\bar z\bar z\bar z}
\partial_\mu \bar z \cdot \l^{i\zb}\right)\nonumber \\
&&
+\ft 12 g^{z\zb}\D_z Z_z\cdot g^{z\zb}\left(\ft 12 \ve^{ik}\ve^{jl}
\<\bar U_{\zb},\tilde Y_{ij}\>\<\bar U_{\zb}, \tilde Y_{kl}\>
- \<\bar U_{\zb},\tilde \F^-_{ab}\>\<\bar U_{\zb}, \tilde \F^{-ab}\>\right)
\nonumber \\
&&-\ft {1}{2}\rmi\ve^{ik}\ve^{jl}\left(\D_z Z_z\cdot\<\bar V,\tilde Y^{ij}\>
+(-C_{zzz} + \D_z\D_z Z_z)\, g^{z\zb}\<\bar U_{\zb},\tilde Y_{ij}\>\right)
\bar\l^z_k\l^z_l
\nonumber  \\
&& -2\rmi g^{z\zb}\D_z Z_z\cdot\ve^{ij}\bar\l^z_i\<\bar U_{\zb},\tilde\Lambda_j\>
\nonumber\\ &&
-\ft {1}{12}\left( \D_z\D_z\D_z Z_z -2\D_z C_{zzz}\right)
\ve^{ij}(\bar\l^z_i\s^{ab}\l^z_j)\,\ve^{kl}(\bar\l^z_k\s_{ab}\l^z_l)  \nonumber \\
&&-\ft 12\rmi\left(\D_z Z_z\cdot\<\bar V,\tilde \F^-_{ab}\>
+(-C_{zzz} + \D_z\D_z Z_z)\, g^{z\zb}\<\bar U_{\zb},\tilde \F^-_{ab}\>\right)
\ve^{kl}(\bar\l^z_k\s^{ab}\l^z_l)\,, \nonumber \\
0 &=& -2\rmi g'_{z\bar z}\Big[ e^{-1}\partial_\mu (e \nabla^\mu \bar z)
+2\rmi({\cal Q}_\mu - A_\mu)\nabla^\mu\bar z\nonumber\\ &&
+\ft 12\rmi({\cal Q}_\mu - A_\mu)(\bar\psi_i^\mu\l^{i\bar z})
+2{\bar \psi}^{i[\mu}\g_\mu\psi_i^{\nu]}\nabla_\nu {\bar z}\nonumber \\
&& +3\bar \chi_i\l^{i\zb}
   -\ft 12 \bar\l^{i\zb} \g_\mu \phi_i^\mu
   + \Gamma_{\zb\zb}^{\zb} \partial_\mu \zb\cdot\nabla^\mu{\zb}
   +\ft 18 \bar\psi^i_\mu\g^\mu\s\cdot T_{ij}\l^{j\zb}\nonumber \\
&&-\bar \psi_i^\mu\left(\nabla_\mu \l^{i\zb}
+\ft 12\rmi({\cal Q}_\mu-A_\mu)\l^{i\zb} +\ft 12\rmi g^{z\zb}\<U_z,\tilde Y^{ij}
+\s\cdot\tilde\F^+\ve^{ij}\>\psi_{\mu j}\right)\Big]
\nonumber \\
&&+2\rmi Z_z\left(\partial_\mu\zb\cdot\D_{\zb} \bar Z_{\bar z}
(\nabla^\mu\zb -\bar\psi^\mu_i\l^{i\zb}) +\bar\psi^\mu_i\<\bar V,\tilde Y^{ij}
+\s\cdot\tilde \F^+\ve^{ij}\>\psi_{\mu j}\right)\nonumber \\
&&+\ft 14 C_{zzz}g^{z\zb}g^{z\zb}\ve^{ik}\ve^{jl}\<\bar U_{\zb},\tilde Y_{ij}\>
\<\bar U_{\zb},\tilde Y_{kl}\>
-\ft 12 C_{zzz}g^{z\zb}g^{z\zb}\<\bar U_{\zb},\tilde \F^-_{ab}\>\<\bar U_{\zb},
\tilde \F^{-ab}\> \nonumber \\
&&-\ft 12\rmi \ve^{ik}\ve^{jl}\left(C_{zzz}\<\bar V,\tilde Y_{ij}\> +\D_z C_{zzz}
\cdot g^{z\zb}\<\bar U_{\zb}, \tilde Y_{ij}\>\right)\bar\l^z_k\l^z_l
+\ft 14 \<U'_z, \tilde \F^+_{\mu\nu}\>  T^{\mu\nu}_{ij}\ve^{ij}\nonumber \\
&&+\ft 12\rmi \left(C_{zzz}\<\bar V,\tilde \F^-_{ab}\>
+\D_z C_{zzz}\cdot g^{z\zb}\<\bar U_{\zb}, \tilde \F^-_{ab}\>\right)\,
\ve^{ij}(\bar\l^z_i\s^{ab}\l^z_j) \nonumber \\
&&+\ft 1{12} \left(\D_z \D_z C_{zzz} +\ft 12 {\cal O}_{zzzzz}\right)
\ve^{ij}(\bar\l^z_i\s_{ab}\l^z_j) \ve^{kl}(\bar\l^z_k\s_{ab}\l^z_l) \nonumber \\
&& -2\rmi C_{zzz}g^{z\zb}\ve^{ij}\bar\l^z_i\<\bar U_{\zb},\tilde\Lambda_j\>\,.
\label{eomz}
\err
The metric in front of the kinetic term of the scalar in the fourth equation is
positive because of the physical condition \eqn{physicalcond}.
Again, all these equations reduce to the equations of section \ref{s:genBI} if
$Z_z=0$ and the same conclusions can be drawn as in section
\ref{s:genBI}. Therefore, we conclude at this point that the `special
case' is a valid alternative for a theory with $N=2$ supergravity
and one vector multiplet.

\section{Conclusions}\label{ss:concl}

We have presented a fully symplectic invariant formulation of the
coupling of an arbitrary number $n$ of vector multiplets to $N=2$
supergravity in 4 dimensions by using superconformal tensor calculus.
This approach does not start from a prepotential, but rather from a
$2(n+1)$ symplectic vector of chiral superconformal multiplets.
We imposed the reducibility
constraints \eqn{bian1}--\eqn{bian4} of chiral multiplets in
supergravity to end up with vector multiplets in a superconformal
background. Furthermore, we imposed the symplectic covariant
defining equations of special \Ka\ geometry, and supersymmetric
partners thereof. The bosonic defining equations include a breaking
of dilatations and $U(1)$-transformations, the special conformal
symmetry being broken as usual by imposing a
constraint \eqn{EC} on the dilatational gauge field. In the fermionic
sector, one of the constraints, \eqn{Sgauge}, breaks special
supersymmetry. This results in unbroken Poincar\' e supersymmetry and $SU(2)$ gauge
symmetry. The other constraints are determined by demanding the
preservation of the \Poin\ supersymmetry.
\par
The combination of all the special \Ka\ constraints
\eqn{VVbar}--\eqn{VUal} and their supersymmetric partners with the
generalized Bianchi identities of the chiral multiplets gave rise to
a full set of field equations for the vector multiplet fields.
\par
Furthermore, we also discussed part of the equations of motion for
the gravity sector. This could be done by imposing a new symplectic
constraint \eqn{FET} and its supersymmetric partners. The full
analysis would need a second compensating multiplet.
\par
Finally, we did the same analysis for a weaker definition of the
special \Ka\ constraints where the constraint \eqn{VUal} is
replaced by \eqn{UU}. This is a weaker constraint for the case of one
physical vector multiplet. In appendix~C of \cite{CRTVP} two examples
were given where (\ref{VUal}) is not satisfied. They are not suitable
for illustrating non-trivial aspects of our construction. The first
example does not fulfil the positivity condition
(\ref{physicalcond}). The second example does agree
with our definition, but is trivial in the sense that there $\D_z
U_z= 0$. Therefore, the extra terms that appear in this paper are
absent for this model.
A nontrivial realization of our new models can e.g.\ be obtained by
taking
\begin{equation}
V= \left(\begin{array}{c}
1\\ az\\ -z^3 \\ 3z^2
\end{array}\right)e^{K/2}\,.
\end{equation}
For $a=1$ it is the well-known $SU(1,1)/U(1)$ symmetric space with
positive metric in the complex upper half plane. Deviating from this
value gives a non-zero value to
\begin{equation}
\langle V, U_z \rangle = \frac{3i(a-1)z^2}{z^3 - 3az^2{\bar z} +3az{\bar z}^2
  -{\bar z}^3}\,.
\end{equation}
Then the new metric
\begin{equation}
g'_{z{\bar z}}=\frac{-3a(z-\zb)^2}{(z^2 + (1-3a)z\zb + \zb^2)^2}
\end{equation}
has a well-defined positivity domain, and
\begin{equation}
  C_{zzz}  =  \frac{6ia(z -\zb)}{(z^2 + (1 - 3a)z\zb + \zb^2)^2}
\label{Czzznew}
\end{equation}
is not covariantly constant.
\medskip

\section*{Acknowledgments.}

\noindent
This work was
supported by the European Commission TMR programme
ERBFMRX-CT96-0045.
\newpage

\appendix
\setcounter{equation}{0}
\section{A basis for symplectic vectors\label{s:basis}}
In this appendix we show that
\bq
{\cal W} = (V,U_\a, \bar V, \bar U_{\bar \a})
\eq
is a basis for symplectic vectors. Since we are dealing with a
$2(n+1)$-dimensional vector space we only have to show that these vectors are
independent.\\
\noindent
{\sl Proof:}
Suppose
\bq
\l^0 V+ \l^{\bar 0} \bar V + \l^\a U_\a  + \l^{\bar \a} \bar U_{\bar \a} = 0\,,
\eq
then it follows that all $\l^i=0$ if and only if the determinant
obtained by left symplectic inner products with, respectively, $\bar
V$, $V$, $\bar U_{\bar \beta}$ and $U_\beta$, is non--zero:
\bq
\det \left(\begin{array}{cccc}
-\rmi&0&0&\<  \bar V, \bar U_{\bar \a} \>  \\
0&\rmi&\< V,U_\a\> &0\\
0&\<\bar U_{\bar \beta}, \bar V\> & \rmi g_{\alpha\bar \beta}&0\\
\< U_\beta ,V\> &0&0&-\rmi g_{\a\bar\b}\\
\end{array}\right ) \ne 0\label{determinant}\,.
\eq
We can split this up in two cases:\\
\noindent
$1.\, \underline{{\rm The\, generic\, case:}}$\\
Then $\< V, U_\a\>  = 0$, and (\ref{determinant}) is
\bq
(\det g_{\a\bar\b})^2 > 0\,,
\eq
which is satisfied by the metric.
\noindent \\
$2.\, \underline{{\rm The\, special\, case:}}$\\
Then we define $Z_z=\< V,U_z\> $ and the determinant equation leads to
\bq
(g_{z\bar z} - Z_z\bar Z_{\bar z})^2 \ne 0\,.
\eq
However, this follows from the `physical' condition on the
sections that leads to the right signs for the kinetic energy of the scalars
and the vectors, cf \eqn{physicalcond}.
\par
Now that we have a basis, we can expand every symplectic vector in this basis.
Take a generic symplectic vector $X_A$, where the index $A$ denotes a
generic index. It is again useful to separate two cases.\\
\noindent
$1.\, \underline{{\rm The\, generic\, case:}}$\\
This leads to
\brr
X_A&=&\rmi \< \bar V, X_A\>  V - \rmi\< V,X_A\>  \bar V\nonumber\\
&&+ \rmi g^{\a\bar\a}\left(
\< U_\a,X_A\>  \bar U_{\bar \a}-\< \bar U_{\bar \a}, X_A\>  U_\a\right) \,.
\label{expn>1}
\err
\noindent
$2.\, \underline{{\rm The\, special\, case:}}$\\
In the basis ${\cal W}$, the expansion becomes
\brr\label{expn1}
X_A&=& -\rmi g'^{z\zb}\bpl (-g_{z\bar z} \< \bar V, X_A\>
+ \rmi \bar Z_{\bar z} \< U_z,X_A\>  ) V\nonumber\\
&&+ (g_{z\bar z} \< V, X_A\>  + \rmi Z_z \< \bar U_{\bar z} , X_A\>  ) \bar
V\nonumber\\
&&+ (-\rmi \bar Z_{\bar z} \< V,X_A\>  + \< \bar U_{\bar z} , X_A\>  ) U_z\nonumber\\
&& - (\rmi Z_z\<  \bar V, X_A\>  + \< U_z, X_A\>  ) \bar U_{\bar z}\bpr\, .
\err
In this case we better use the basis
\begin{equation}
{\cal W}' = (V,U'_z, \bar V, \bar U'_{\bar z})\,.    \label{expn2}
\end{equation}
The same formulae hold as above, when replacing $g_{\alpha\bar
\beta}$ with $g'_{z\bar z}$.

\end{document}